\documentclass[pra,aps,superscriptaddress,twocolumn,letter,nopacs,nofootinbib,longbibliography]{revtex4-2}

\usepackage[english]{babel}
\usepackage[utf8x]{inputenc}
\usepackage[T1]{fontenc}

\usepackage{amsthm}
\theoremstyle{plain} 
\usepackage{graphicx}
\usepackage[colorinlistoftodos]{todonotes}
\usepackage[colorlinks=true, allcolors=blue]{hyperref}
\usepackage{braket}
\usepackage{comment}
\usepackage{float}
\usepackage{bbm,amsfonts}
\usepackage[shortlabels]{enumitem}
\usepackage{hyperref}
\usepackage{amsthm}
\usepackage{xcolor}
\usepackage{mathtools}
\usepackage[title]{appendix}

\hypersetup{
	colorlinks=true,  
	linkcolor=blue,   
	citecolor=blue,   
	urlcolor=blue     
}

\def\>{\rangle}
\def\<{\langle}

\newcommand{\abs}[1]{\left| {#1} \right|}

\newtheorem{theorem}{Theorem}
\newtheorem{example}[theorem]{Example}
\newtheorem*{example*}{Example}

\newtheorem{definition}[theorem]{Definition}
\newtheorem{remark}[theorem]{Remark}
\newtheorem*{remark*}{Remark}
\newtheorem{corollary}[theorem]{Corollary}
\newtheorem{conjecture}[theorem]{Conjecture}
\newtheorem{proposition}[theorem]{Proposition}

\usepackage{bbold}
\usepackage{xspace}
\usepackage{array}

\usepackage{mathtools}

\usepackage{float}

\DeclareMathOperator{\co}{c}
\DeclareMathOperator{\s}{s}
\usepackage{graphicx}               
\usepackage{amsmath,amssymb}

\begin{document}


	
\title{Genuinely quantum SudoQ and its cardinality}

\author{Jerzy Paczos}
\affiliation{Faculty of Physics, University of Warsaw, ul. Pasteura 5, 02-093 Warszawa, Poland}
\author{Marcin Wierzbiński}
\affiliation{Faculty of Mathematics, Informatics and Mechanics, University of Warsaw, ul. Banacha 2, 02-097 Warszawa, Poland}
\author{Grzegorz Rajchel-Mieldzio{\'c}}
\affiliation{Center for Theoretical Physics, Polish Academy of Sciences, Al. Lotnik\'ow 32/46, 02-668 Warszawa, Poland}
\author{Adam Burchardt}
\affiliation{Institute of Theoretical Physics, Jagiellonian University, ul. {\L}ojasiewicza 11, 30–348 Krak\'ow, Poland}
\author{Karol {\.Z}yczkowski}
\affiliation{Center for Theoretical Physics, Polish Academy of Sciences, Al. Lotnik\'ow 32/46, 02-668 Warszawa, Poland}
\affiliation{Institute of Theoretical Physics, Jagiellonian University, ul. {\L}ojasiewicza 11, 30–348 Krak\'ow, Poland}




\date{June 1, 2021}

\begin{abstract}
We expand the quantum variant of the popular game Sudoku by introducing the notion of \emph{cardinality} of a quantum Sudoku (SudoQ), equal to the number of distinct vectors appearing in the pattern. Our considerations are focused on the \emph{genuinely quantum solutions} – the solutions of size $N^2$ that have cardinality greater than $N^2$, and therefore cannot be reduced to classical counterparts by a unitary transformation. We find the complete parameterization of the genuinely quantum solutions of $4\times 4$ SudoQ game and establish that in this case the admissible cardinalities are 4, 6, 8 and 16. In particular, a solution with the maximal cardinality equal to $16$ is presented. 
Furthermore, the parametrization enabled us to prove a recent conjecture of Nechita and Pillet for this special dimension.
In general, we proved that for any $N$  it is possible to find an $N^2\times N^2$ SudoQ solution of cardinality $N^4$, which for a prime $N$ is related to a set of $N$ mutually unbiased bases of size $N^2$. 
Such a construction of $N^4$ different vectors of size $N$ yields a set of $N^3$ orthogonal measurements.
 \end{abstract}

\maketitle

\section{Introduction}

Numerous classical concepts have their quantum counterparts.
It is no different with \textit{Latin squares} \cite{fisher_yates_1934}, which are $N\times N$ arrays of $N$ distinct symbols, arranged so that no two elements are the same in any row or column.
Extension of these to \emph{quantum Latin squares} (QLS) was proposed in \cite{musto2016quantum}. These basic structures in quantum information play a central role in quantum teleportation \cite{PhysRevLett.70.1895}, dense coding \cite{Gu_2012} and other quantum information protocols \cite{Musto_2017}. Moreover, two QLS can be entangled in such a way that they cannot be expressed as two separated arrangements \cite{PhysRevA.97.062326}.
One of the recent generalizations of QLS are block bistochastic matrices \cite{Benoist_2017}, later studied in \cite{gemma_2020} under the name of quantum magic squares, in which the same elements may appear in the same row or line.

For any Latin square it is possible to create a QLS by association to each number $i=1,\ldots, N$ a vector from the computational basis, i.e. $i \mapsto \ket{i}$.
QLS which can be obtained by such connection are called \emph{classical}.
On the other hand, there is no canonical way to associate a Latin square to a generic quantum one.
Therefore, QLS provide richer structure than Latin squares; however, not all of them are truly quantum, i.e. a variety of them can be transformed into classical QLS by a unitary action on all vectors, what was first noted by Musto and Vicary \cite{Musto_2019}. 
We call such QLS \emph{apparently quantum} and, to quantify the contrast between various classes of QLS we introduce the notion of cardinality -- the number of its vectors different up to a phase.
QLS of cardinality larger than $N$ cannot be unitarily transformed into a classical solution, which is why they can be called \emph{genuinely quantum}.

The special kind of Latin squares are Sudoku, the designs which admit additional constraints that also no two elements repeat in any out of $N$ disjoint blocks.
Following the idea of quantum Latin squares, Nechita and Pillet introduced a quantum version of Sudoku \cite{Nechita2020SudoQA}, which was named \emph{SudoQ} by adding orthogonality constraints on each row, column and block on a given array of vectors.
While the number of possible Sudoku patterns of size nine is finite (and equal to 6,670,903,752,021,072,936,960 -- see~\cite{enumerating_sudoku,count}), the number of different SudoQ designs is infinite by construction.
Since SudoQ designs form a subset of QLS, the definition of cardinality holds without any modifications.

The aim of this paper is to study cardinality of SudoQ, as well as to explore its relationship to other quantum mechanical notions, such as mutually unbiased bases.
Two bases are said to be \textit{unbiased} if for a system prepared in an eigenstate of one base, all outcomes of the measurement with respect to the other basis are predicted to occur with equal probability \cite{Bengtsson_2007}. 
\textit{Mutually unbiased bases} (MUBs) are useful in various quantum informational tasks: in quantum state tomography \cite{WOOTTERS19871}, or quantum cryptography \cite{PhysRevLett.88.127902}, among many others. 
The number of MUBs in dimension $d$, cannot exceed $d+1$.
A \textit{complete} set of $d+1$ MUBs has been constructed for all prime power dimensions \cite{MUBalg}, for instance with the \textit{Heisenberg-Weyl method} \cite{MUBs25,hiesmayr2021detecting}. 
Vectors from a complete set of MUBs form a \textit{projective $2$-design} \cite{Welch,klappenecker2005mutually}, which means that they faithfully approximate the state space for any degree 2 polynomial in state coefficients. 


To further motivate importance of the study of quantum combinatorial designs, notice that vectors from each row, column, or block in such SudoQ grid form a von Neumann measurement. 
Therefore, by considering SudoQ, we address the question concerning existence of von Neumann measurements in the Hilbert space of dimension $N^2$, which are determined by $N^4$ pure states.
Any SudoQ design provides $3N^2$ orthogonal measurements, related to columns, rows and blocks of the SudoQ design.



This paper is organized as follows. 
In section \ref{sec:preliminaries} we introduce definition of cardinality for quantum Latin squares along with apparently and genuinely quantum solutions of QLS. 
 Section \ref{sec:quantum_sudoku} dwells on quantum Sudoku as a special case of QLS. In subsection \ref{sec:param:4x4}, we provide full characterization of cardinality for $4\times 4$ SudoQ, along with the parametrization of the solutions of the $4\times4$ SudoQ.
Furthermore, in section \ref{subsec:9x9} we consider general $N^2 \times N^2$ grids and characterize properties of their quantum solutions, especially those of the maximal cardinality $N^4$.
Section \ref{MUBs} presents the standard Heisenberg-Weyl method of constructing MUBs for prime power dimensions, as well as adaptation of this technique in construction of SudoQ grids. 
Besides, in Section \ref{SudoqqCubes}, we show how to generalize this approach for \emph{SudoQ cubes} and \emph{hypercubes}. 
Similarly to MUBs, vectors from SudoQ grid form complex projective design. 
Finally, section \ref{comparison} compares both designs against the Welch bound. 
Moreover, we investigate the \textit{local structure} of SudoQ vectors belonging to the composed space $\mathcal{H}_N \otimes \mathcal{H}_N$ .

\section{Quantum Latin squares}
\label{sec:preliminaries}
The domain of quantum designs has its roots in the classical combinatorial designs called Latin squares~\cite{fisher_yates_1934}.
\begin{definition}\label{def:latin_squares}
A Latin square (LS) is an $N \times N$ array of $N$ distinct elements, arranged so that no element repeats in any row or column.
\end{definition}
\begin{example*}\emph{Latin square of size $N=3$.}
\begin{equation}\label{eq:LS3}
    \begin{tabular}{!{\vrule width 1pt}c|c|c!{\vrule width 1pt}}
    \noalign{\hrule height 1pt}
    $\quad$ 1 $\quad$ & $\quad$ 2 $\quad$ & $\quad$ 3 $\quad$ \tabularnewline
    \hline
    3 & 1 & 2  \tabularnewline
    \hline
    2 & 3 &1  \tabularnewline
    \noalign{\hrule height 1pt}
    \end{tabular}    
\end{equation}
\end{example*}

The main goal of the paper is to investigate a special case of quantum Latin squares~\cite{musto2016quantum,PhysRevA.97.062326}; therefore, let us revoke their definition.
\begin{definition}\label{def:qls}
A quantum Latin square (QLS) is an $N\times N$ array of vectors from $N$-dimensional Hilbert space $\mathcal{H}_N$, arranged in such a way that each row and column forms an orthonormal basis of $\mathcal{H}_N$.
\end{definition}
\begin{example*}\emph{Quantum Latin square of size $N=3$.}
\begin{equation}\label{app_quant_ex}
    \begin{tabular}{!{\vrule width 1pt}c|c|c!{\vrule width 1pt}}
    \noalign{\hrule height 1pt}
    $\ket{1}$ & $\frac{1}{\sqrt{2}}\left(\ket{2}+\ket{3}\right)$ & $\frac{1}{\sqrt{2}}\left(\ket{2}-\ket{3}\right)$ \tabularnewline
    \hline
    $\frac{1}{\sqrt{2}}\left(\ket{2}-\ket{3}\right)$ & $\ket{1}$ & $\frac{1}{\sqrt{2}}\left(\ket{2}+\ket{3}\right)$  \tabularnewline
    \hline
    $\frac{1}{\sqrt{2}}\left(\ket{2}+\ket{3}\right)$ & $\frac{1}{\sqrt{2}}\left(\ket{2}-\ket{3}\right)$ & $\ket{1}$  \tabularnewline
    \noalign{\hrule height 1pt}
    \end{tabular}
\end{equation}
\end{example*}
In this example vectors $$\{ \ket{1}, \frac{1}{\sqrt{2}}\left(\ket{2}+\ket{3}\right), \frac{1}{\sqrt{2}}\left(\ket{2}-\ket{3}\right)\}$$ form an orthonormal basis of a 3-dimensional Hilbert space.

QLS form richer structure than Latin squares; thus, it is useful to designate those quantum Latin squares which have their classical counterparts.

\begin{definition}\label{def:classical_QLS}
A quantum Latin square which consists of elements only from the computational basis \mbox{$\mathcal{B} = \{\ket{1},...,\ket{N}\}$} is called \emph{classical}.
\end{definition}

Equivalence of classical QLS and Latin squares can be visualized by treating every element of QLS as a symbol from Def.~\ref{def:latin_squares}.
Let us note the remark which will be crucial for our paper, since it motivates our investigation of QLS and quantum Sudoku.

\begin{proposition}\label{proposition:u_similar}
Any $N\times N$ QLS of $N$ distinct entries is unitarily similar to a classical one (with elements from computational basis only). Such quantum Latin square will be called \mbox{\emph{apparently quantum}}.
\end{proposition}
\begin{proof}
Validity of the above statement follows from the existence of a unitary operator which transforms any orthonormal basis to the computational one.
\end{proof}

In particular, the QLS (\ref{app_quant_ex}) is apparently quantum, since it contains only 3 distinct elements, so it can be transformed to the classical solution~(\ref{eq:LS3}) by an orthogonal rotation in the subspace spanned by $\ket{2}$ and $ \ket{3}$.

Mitigated by the aforementioned proposition we are able to introduce a new notion, which is of central importance to our results since it encompasses the core distinction between the classical and the quantum case.

\begin{definition}\label{def:cardinality}
The cardinality $c$ of a QLS is the number of its vectors distinct up to a global phase.
\end{definition}

Any classical LS is characterized by the smallest cardinality $c=N$, while for any QLS its maximal cardinality is $c_{max} = N^2$.

Usefulness of the new notion stems from its invariance under unitary operations.
This way it can be used to study non-classicality of a given QLS since, similarly to entanglement measures, it is not affected by free operations (unitaries as opposed to the local unitaries in the case of local operations and classical communication).

Analogously to Prop. \ref{proposition:u_similar} we are able to prove the following statement.

\begin{proposition}\label{prop:genuinely_quantum_QLS}
Any $N\times N$ QLS with the cardinality greater than $N$ is not unitarily similar to a classical one. Such QLS will be called \mbox{\emph{genuinely quantum}}.
\end{proposition}

Note that this remark could be used to simplify the proof of Proposition 13 on page 6 in Ref.~\cite{Musto_2019}.

In the subsequent parts of this paper we will denote vectors that are linear combinations of vectors from computational basis $\ket{i}$ and $\ket{j}$ as $\ket{a_{ij}},\ket{b_{ij}}$, where coefficients of these combinations may be equal to zero.


\begin{remark*}
There are no genuinely quantum $2\times 2$ QLS.
\end{remark*}
\begin{proof}
By contradiction, if there were more than 2 distinct vectors in a $2\times 2$ lattice then one of them must be orthogonal to the two others, which is not possible in a 2-dimensional Hilbert space.
\end{proof}

\begin{remark*}
There are no genuinely quantum $3\times 3$ QLS.
\end{remark*}
\begin{proof}
Without loss of generality let us consider a 3$\times$3 grid with vectors from the computational basis $\mathcal{B}$ in the first row and unknown vectors in the other two rows
\begin{equation}
\begin{tabular}{!{\vrule width 1pt}c!{\vrule width 1pt}c!{\vrule width 1pt}c!{\vrule width 1pt}}
    \noalign{\hrule height 1pt}
    $\ket{1}$ & $\ket{2}$ & $\ket{3}$ \tabularnewline
    \noalign{\hrule height 1pt}
    $\ket{a_{23}}$ & $\ket{a_{13}}$ & $\ket{a_{12}}$ \tabularnewline
    \noalign{\hrule height 1pt}
    $\ket{b_{23}}$ & $\ket{b_{13}}$ & $\ket{b_{12}}$ \tabularnewline
    \noalign{\hrule height 1pt}
\end{tabular}\,\, .    
\end{equation}
Observe that choosing one of the vectors in the second row to belong to the computational basis $\mathcal{B}$ forces the others to be such (as the scalar products of each two vectors in each row must be zero). On the other hand, if we set all the vectors in the second row not to belong to this basis, then they cannot be mutually orthogonal. Therefore, they must be all computational. The same holds for the vectors in the third row.
\end{proof}
In the rest of the paper we will deal with the special type of QLS, namely the quantum Sudoku.

\section{Quantum Sudoku}\label{sec:quantum_sudoku}
Motivated by paper~\cite{Nechita2020SudoQA} we extend Definition~\ref{def:qls} introducing the quantum Sudoku as a special case of QLS.
\begin{definition}
A quantum Sudoku (SudoQ) is an $N^2 \times N^2$ QLS with additional constraints that each of the disjoint blocks of size $N$ forms an orthonormal basis.
\end{definition}

Let us proceed to investigate instances of partially filled SudoQ, with the particular emphasis on its solvability.

\begin{definition}
A quantum grid is a SudoQ grid whose some entries can be blank.
\end{definition}
\begin{example*}\emph{Quantum grid of size $2^{2} \times 2^{2}$.}
\begin{equation}
    \begin{tabular}{!{\vrule width 1pt}c|c!{\vrule width 1pt}c|c!{\vrule width 1pt}}
    \noalign{\hrule height 1pt}
    $\ket{1}$ & $\ket{2}$ & $\ket{3}$ & $\ket{4}$ \tabularnewline
    \hline
    $\frac{1}{\sqrt{2}}\left(\ket{1}+\ket{2}\right)$ & $\frac{1}{\sqrt{2}}\left(\ket{1}-\ket{2}\right)$ & . & . \tabularnewline
    \noalign{\hrule height 1pt}
    . & . & . & . \tabularnewline
    \hline
    . & . & . & . \tabularnewline
    \noalign{\hrule height 1pt}
    \end{tabular}
\end{equation}
\end{example*}

We are particularly interested in a subclass of those grids which can be filled by vectors without breaking the orthogonality conditions. 

\begin{definition}
A quantum grid is called solvable if it has at least one solution.
\end{definition}
\begin{example*}\emph{Solvable quantum grid with 6 clues.}
\begin{equation}
    \begin{tabular}{!{\vrule width 1pt}c|c!{\vrule width 1pt}c|c!{\vrule width 1pt}}
    \noalign{\hrule height 1pt}
    $\ket{1}$ & $\ket{2}$ & $\ket{3}$ & $\ket{4}$ \tabularnewline
    \hline
    . & . & $\frac{1}{\sqrt{2}}\left(\ket{1}+\ket{2}\right)$ & $\frac{1}{\sqrt{2}}\left(\ket{1}-\ket{2}\right)$ \tabularnewline
    \noalign{\hrule height 1pt}
    . & . & . & . \tabularnewline
    \hline
    . & . & . & . \tabularnewline
    \noalign{\hrule height 1pt}
    \end{tabular}
\end{equation}
\end{example*}

One of the combinatorial problems concerning Sudoku is the minimal number of clues admitting a unique solution; therefore, we introduce a similar concept for its quantum counterpart.

\begin{definition}
A quantum grid is called uniquely solvable if it has a unique solution.
\end{definition}
\begin{example*}\emph{Uniquely solvable quantum grid of size $2^{2}\times 2^{2}$.}
\begin{equation}
    \begin{tabular}{!{\vrule width 1pt}c|c!{\vrule width 1pt}c|c!{\vrule width 1pt}}
    \noalign{\hrule height 1pt}
    $\ket{1}$ & $\ket{2}$ & $\ket{3}$ & $\ket{4}$ \tabularnewline
    \hline
    $\ket{3}$ & $\ket{4}$ & $\frac{1}{\sqrt{2}}\left(\ket{1}+\ket{2}\right)$ & $\frac{1}{\sqrt{2}}\left(\ket{1}-\ket{2}\right)$ \tabularnewline
    \noalign{\hrule height 1pt}
    $\ket{2}$ & $\ket{1}$ & . & . \tabularnewline
    \hline
    . & . & . & . \tabularnewline
    \noalign{\hrule height 1pt}
    \end{tabular}
\end{equation}
\end{example*}


Since SudoQ is a restriction of the QLS, by analogy we introduce \emph{cardinality} of a SudoQ, i.e. the number of its distinct entries (up to a phase), along with \emph{classical}, \emph{apparently quantum} and \emph{genuinely quantum} solutions, see Def.~\ref{def:classical_QLS} and \ref{def:cardinality} along with Prop.~\ref{proposition:u_similar} and \ref{prop:genuinely_quantum_QLS}.

\subsection{Quantum Sudoku $4\times 4$}\label{sec:param:4x4}

In this section we present the main results of analysis of $4\times 4$ SudoQ, that is the list of all possible cardinalities of such a design, as well as its parametrization.

\begin{theorem}\label{cardinalities}
The only admissible cardinalities of a $4\times 4$ SudoQ read $c=4, 6, 8$ and $16$.
\end{theorem}
Proof of the above theorem is moved to Appendix~\ref{sec:proof_cardinalities}.

Subsequently, we provide the reader with the parametrization of the SudoQ solutions with the maximal cardinality $c=16$. We present the derivation of this parametrization, and show how to obtain a parametrization of solutions with $c=6$ and $c=8$ in Appendix~\ref{sec:16parametrization}.
\mbox{}\\\mbox{}\\
The general form of the solution of a 4$\times$4 SudoQ is the following
\begin{equation}\label{generalsolution}
    \begin{tabular}{!{\vrule width 1pt}c|c!{\vrule width 1pt}c|c!{\vrule width 1pt}}
    \noalign{\hrule height 1pt}
    $e_1$ & $e_2$ & $f_1$ & $f_2$ \tabularnewline
    \hline
    $e_3$ & $e_4$ & $f_3$ & $f_4$ \tabularnewline
    \noalign{\hrule height 1pt}
    $v_1$ & $v_2$ & $u_1$ & $u_2$ \tabularnewline
    \hline
    $v_3$ & $v_4$ & $u_3$ & $u_4$ \tabularnewline
    \noalign{\hrule height 1pt}
    \end{tabular}\ ,
\end{equation}
where $\{e_i\}$, $\{f_i\}$, $\{v_i\}$ and $\{u_i\}$ are the orthonormal bases of the four dimensional Hilbert space $\mathcal{H}_4$. Let us set $\{e_i\}$ to form the computational basis $\mathcal{B} = \{\ket{i}\}$ and express other vectors in terms of $e_i$. It will be helpful to define unitary matrices $U_{ef}$, $U_{ev}$, $U_{eu}$, such that
\begin{equation}\label{matrixdef}
U_{ef}\ket{e_i}=\ket{f_i},\qquad U_{ev}\ket{e_i}=\ket{v_i}, \qquad U_{eu}\ket{e_i}=\ket{u_i}.
\end{equation}
Then columns of the matrices $U_{ef}$, $U_{ev}$, $U_{eu}$ are vectors from $\{f_i\}$, $\{v_i\}$ and $\{u_i\}$ respectively, expressed in terms of the computational basis.
If we consider the SudoQ with $c=16$, then

\begin{equation}
\begin{split}
&\text{\small $
U_{ef}=
\begin{bmatrix}
0 & 0 & \co_\alpha & \s_\alpha \\
0 & 0  & e^{i\phi}\s_\alpha & -e^{i\phi}\co_\alpha \\
\co_\alpha & \s_\alpha  & 0  & 0 \\
e^{i\varphi}\s_\alpha& -e^{i\varphi}\co_\alpha & 0 & 0
\end{bmatrix},$}\\ 
\\
&\text{\small $
U_{ev}=
\begin{bmatrix}
0 & \co_\gamma & 0 & \s_\gamma \\
\co_\gamma & 0  & \s_\gamma & 0 \\
0  & e^{i\zeta}\s_\gamma & 0  & -e^{i\zeta}\co_\gamma \\
e^{i\eta}\s_\gamma & 0  & -e^{i\eta}\co_\gamma & 0
\end{bmatrix},$}\\ 
\\
&\text{\scriptsize $
U_{eu}=
    \begin{bmatrix}                      
    \s_\alpha \s_\gamma&  \co_\alpha \s_\gamma &
 \s_\alpha \co_\gamma & \co_\alpha \co_\gamma \\
    -e^{i\phi}\co_\alpha \s_\gamma &
    e^{i\phi}\s_\alpha \s_\gamma &
    -e^{i\phi}\co_\alpha \co_\gamma &
    e^{i\phi}\s_\alpha \co_\gamma\\
    -e^{i\zeta}\s_\alpha\co_\gamma &
    -e^{i(\zeta)}\co_\alpha \co_\gamma &
    e^{i\zeta}\s_\alpha \s_\gamma  &
    e^{i\zeta}\co_\alpha \s_\gamma \\
    e^{i(\phi+\eta)}\co_\alpha \co_\gamma &
    -e^{i(\phi+\eta)}\s_\alpha \co_\gamma &
    -e^{i(\phi+\eta)}\co_\alpha \s_\gamma &
    e^{i(\phi+\eta)}\s_\alpha \s_\gamma
    \end{bmatrix}$}
\end{split}
\label{parametrisation}
\end{equation}
where $s_\alpha$ denotes $\sin{\alpha}$ and $c_\alpha$ denotes $\cos{\alpha}$ respectively, while four angle parameters are constrained by $\phi+\eta= \varphi +\zeta$. 
For any values of the parameters Eq.~\ref{parametrisation} gives a valid solution of a SudoQ. Moreover, for a different choice of parameters it yields cardinalities $c=4,8,16$.
As a particularly symmetric construction we provide Example \ref{ex:4x4_card16},
which forms a proper SudoQ of the maximal cardinality, $c=16$.
For clarity we used here non-normalized vectors.
\begin{example}\label{ex:4x4_card16}
{\small
\[
 \begin{tabular}{!{\vrule width 1pt}c|c!{\vrule width 1pt}c|c!{\vrule width 1pt}}
    \noalign{\hrule height 1pt}
    $\ket{1}$ & $\ket{2}$ & $\ket{3}+\ket{4}$ & $\ket{3}-\ket{4}$ \tabularnewline
    \hline
    $\ket{3}$ & $\ket{4}$ & $\ket{1}-\ket{2}$ & $\ket{1}+\ket{2}$ \tabularnewline
    \noalign{\hrule height 1pt}
    $\ket{2}+\ket{4}$ & $\ket{1}-\ket{3}$ & $\ket{1}+\ket{2}+\ket{3}-\ket{4}$ & $\ket{1}-\ket{2}+\ket{3}+\ket{4}$ \tabularnewline
    \hline
    $\ket{2}-\ket{4}$ & $\ket{1}+\ket{3}$ & $\ket{1}+\ket{2}-\ket{3}+\ket{4}$ & $\ket{1}-\ket{2}-\ket{3}-\ket{4}$ \tabularnewline
    \noalign{\hrule height 1pt}
\end{tabular}
\]
}
\end{example}
Above example is especially interesting because it has a maximum degree of superposition, i.e. the entropy of states coefficients in the computational basis.
As shown in Appendix~\ref{sec:shannon}, this solution is distinguished by attaining maximal value of the averaged entropy of components of the vectors.

For any vector $\ket{e_i}$ from the upper left block and any normalized vector $\ket{u_j}$ from the lower right block one has
\[
\abs{\braket{e_i|u_j}}^2=\frac{1}{4}.
\]

This means that bases $\{e_1,\ldots,e_4\}$ and $\{u_1,\ldots,u_4\}$ are mutually unbiased (MUB), i.e. every pair of vectors from both bases admits the same scalar product
\cite{Ivonovic_1981,Wooters_1989}. The same holds true for the pair of bases from the upper right and lower left blocks.
Setting two mutually unbiased bases of order four in either diagonal or anti-diagonal blocks leaves no freedom for the other elements, i.e. it specifies the whole structure uniquely.

As we show in Appendix~\ref{sec:proof_thm}, Example~\ref{ex:4x4_card16} can be solved given only 4 clues, which is also the minimal number in the classical case.

\begin{theorem}\label{thm:4_clues}
For any 4$\times$4 quantum square with $c=16$, the corresponding grid with 4 clues
\begin{equation}
    \begin{tabular}{!{\vrule width 1pt}c|c!{\vrule width 1pt}c|c!{\vrule width 1pt}}
    \noalign{\hrule height 1pt}
    $e_1$ &   &   &   \tabularnewline
    \hline
      &   & $f_3$ &   \tabularnewline
    \noalign{\hrule height 1pt}
      & $v_2$ &   &   \tabularnewline
    \hline
      &   &   & $u_4$ \tabularnewline
    \noalign{\hrule height 1pt}
    \end{tabular}
\end{equation}
is uniquely solvable.
\end{theorem}

 The authors of \cite{Forrow2014ApproachingTM} investigated the issue of minimum number of clues in a classical $4\times 4$ Sudoku and showed that it is equal to $4$. Basing on Theorem~\ref{thm:4_clues} we were able to pose a conjecture.

\begin{conjecture}
Similarly to the classical Sudoku there are no $4\times 4$ SudoQ with 3 clues that are uniquely solvable. 
\end{conjecture}

Using the parametrization presented in Appendix~\ref{sec:16parametrization} we are able to prove in the case of $4\times 4$ SudoQ the conjecture from \cite{Nechita2020SudoQA}, which states that uniquely solvable Sudoku admits only one solution also in the quantum regime.

\begin{proposition}\label{Nechita_conjecture}
Uniquely solvable $4\times 4$ classical Sudoku is also uniquely solvable SudoQ.
\end{proposition}

Proof of the above statement is delegated to Appendix~\ref{sec:conjecture_Nechita}.

\section{General quantum Sudoku}\label{subsec:9x9}
We present a general construction of SudoQ grids in dimension $N^2 \times N^2$, with the particular emphasis on the existence of the solutions of the maximal cardinality. 

Quantum Sudoku is an $N^2 \times N^2$ grid of vectors from a complex projective Hilbert space $\mathcal{H}^{\otimes 2}_N$ which satisfies three orthogonality relations: vectors in each block, row and column form a basis. 
Entries of such a grid might be conveniently labeled by four indices $i,j,k ,\ell =1,\ldots,N$ referring to the $(N(i-1)+ k)$-row and $(N(j-1)+ \ell)$-column in the grid. 
In such a way, indices $i,j$ label consecutive blocks of $N \times N$ vectors in SudoQ grid, while indices 
$k,\ell$ are relevant to the consecutive rows and columns in each block. 
With this notation at hand, we denote by $\ket{v}_{i j k \ell } \in \mathcal{H}^{N^2}$ the vector located on the intersection of $(N(i-1)+ k)$-row with $(N(j-1)+ \ell)$-column in the SudoQ grid. 

\begin{remark}
\label{remark22}
SudoQ is a collection of $N^4$ vectors $ \{ \ket{v}_{i j k \ell } \} \in \mathcal{H}^{N^2}$, such that the marginal sets of $N^2$ vectors $ \{ \ket{v}_{i j k \ell } \} \in \mathcal{H}^{N^2}$ constitutes a basis for any  fixed pairs of indices: $i,j$ or $i, k$ or $j, \ell$. 
\end{remark}

The simplest construction of a classical Sudoku grid is given by the cyclic permutation of rows and columns. 
In such a Sudoku grid, the entry corresponding to $i,j,k,\ell$ coordinates has a value $j+k-1+N(i+\ell-2)\bmod N^2 $. 
We slightly change the notation, and instead of denoting entries as numbers $1,\ldots , N^2$, we shall use pairs of numbers $(p,q)$ where $p, q =1,\ldots,N$.
Therefore, an entry corresponding to indices $i,j,k,\ell$ is given by a pair of numbers $\left( j+ k, i + \ell\right)$.

The standard construction of a Sudoku grid might be canonically identified with the quantum SudoQ with all entries taken from the computational basis $\mathcal{B}$. 
Indeed, consider the following grid of vectors:
\[
 \ket{v}_{i j k \ell } := \ket{j+k} \otimes \ket{i+\ell}\in \mathcal{H}^{\otimes 2}_N
\]
where the addition is considered mod $N$. 
Observe that in this particular type of SudoQ, all vectors $ \ket{v}_{i j k \ell }$ in a given block are determined by two indices $i,j$ form the computational basis. 
Hence, such SudoQ grid has cardinality equal to $N^2$.

In fact, this particular classical Sudoku grid might be generalized to the quantum one in various different ways, where bases of different blocks in general do not overlap.

\begin{proposition}
\label{propp}
Consider two families of $N$ unitary matrices of dimensions $N\times N$, denoted by $\{U_i \}_{i=1}^N $ and $\{V_i \}_{i=1}^N $. 
Let $\ket{u_{s}^{(i)}}$ and $\ket{v_s^{(i)}}$ denote the $s$-th column of the matrices $U_i$ and $V_i$ respectively.
The following set of vectors
\begin{equation}
\ket{w_{ijk\ell}} := \ket{u_{j+k}^{(i)}} \otimes \ket{v_{i+ \ell  }^{(j)}},
\label{allvec}
\end{equation}
where addition is considered modulo $N$, 
constitute the SudoQ grid according to the notation in Remark \ref{remark22}. 
\end{proposition}

\begin{proof}
Using Remark \ref{remark22} we shall verify that vectors $\ket{v_{ijk\ell}} $ constitute a basis of $\mathcal{H}_N^{\otimes 2}$ for any fixed pair of indices $i,j$ and $i, k$ and $j, \ell$. 

Note that for two fixed parameters $i,j$, the vectors $\ket{u_{j+k}^{(i)}}$ are consecutive columns of a unitary operator $U_i$, and hence form a basis in $\mathcal{H}_N$. 
Similarly, the vectors $\ket{v_{i +\ell }^{(j)}}$ form a basis of $\mathcal{H}_N$, thus the tensor products 
$\ket{u_{j+k}^{(i)}} \otimes \ket{v_{i+ \ell  }^{(j)}}$ 
form a basis of $\mathcal{H}_N^{\otimes 2}$. 

Observe that for fixed parameters $i,j,k$, the vectors 
$\ket{u_{j+k}^{(i)}} \otimes \ket{v_{i+ \ell }^{(j)}}$ 
span $N$ dimensional subspace 
$\ket{u_{j+k}^{(i)}} \otimes \mathcal{H}_N$. 
Moreover, for fixed parameters $i,k$ and different values $j=1,\ldots,N$, the vectors 
$\ket{u_{j+k}^{(i)}}$ span the subspace $\mathcal{H}_N^{}$. 
Therefore, for fixed parameters $i,k$, the corresponding vectors $\ket{u_{j+k}^{(i)}} \otimes \ket{v_{i+ \ell }^{(j)}}$ span $\mathcal{H}_N^{\otimes 2}$, and hence form its basis. 
Similar analysis might be performed with respect to fixed indices $j,\ell$.
\end{proof}

\begin{proposition}\label{main:the}
Consider the same family of matrices as in Proposition~\ref{propp}, together with bases formed by treating each column as a vector.
Denote by $c_1$ the number of distinct vectors in the family $\{\mathnormal{U}_i\}_{i=1}^N$ ($c_2$ in the family $\{\mathnormal{V}_i\}_{i=1}^N$ respectively), where equality is considered up to a multiplication by a phase factor. 
Then cardinality of the related SudoQ grid reads $c=c_1 c_2$. 
In particular, if in both families of unitary matrices any column does not occur more than once (up to the multiplication by a phase factor), the corresponding SudoQ grid achieves maximal cardinality. 
\end{proposition}
    \begin{proof}
Observe that the set of vectors $ \ket{v}_{i j k \ell } $ constituting a SudoQ grid according to Eq. (\ref{allvec}) is in fact of the tensor product form:
\[
\{\ket{w}_{i j k \ell }\}_{i,j,k,\ell=1}^N 
=\{ u_{k}^{(i)} \otimes  v_{\ell}^{(j)} \}_{i,j,k,\ell=1}^N 
\]
where $\ket{u_{k}^{(i)}}$ the $k$-th column of $U_i$ operator, and $\ket{v_{\ell}^{(j)}}$ is the $\ell$-th column of $V_j$ respectively. 
Assuming that there are exactly $c_1$ distinct vectors among $u_{k}^{(i)}$ (and $c_2$ among $v_{\ell}^{(j)}$), we conclude the statement.  
    \end{proof}

\begin{corollary}
Notice that the probability that two random unitary matrices 
will share the same column up to the multiplication by a phase factor equals zero. 
Therefore, in any dimension $N$, the SudoQ grid related to two families of random unitary operators $\{U_i \}_{i=1}^N $ is of the maximal cardinality with probability 1.
\end{corollary}

\section{Heisenberg-Weyl SudoQ}
\label{MUBs}

Consider the standard shift and phase operators defined on the $N$-dimensional Hilbert space $\mathcal{H}_N$
\[
X= \sum_{j=1}^N \ket{j}\bra{j+1}, \quad Z=  \sum_{j=1}^N \omega^{j}  \ket{j}\bra{j},
\]
where $\omega= \text{exp} (2 \pi i /N)$ is an $N$-th root of unity.
Addition $\ket{j+1}$ is understood modulo $N$, so these operators can be considered as generalizations of Pauli matrices $\sigma_x$ and $\sigma_z$ corresponding to $N=2$.
Note that vectors constituting eigenbasis of the product $XZ^k$ might be written in a compact way 
\begin{equation}
\ket{j_{k+1}} =\frac{1}{\sqrt{N}} \sum_{\ell=1}^N \omega^{k \ell^2 +j \ell} \ket{\ell},
\label{vectt}
\end{equation}
while the eigenbasis $\{ \ket{j_{1}} \}$ of $Z$ operator is simply the computational basis \cite{hiesmayr2021detecting}. 
For prime dimensions, $N=p$, the eigenvectors of the Heisenberg-Weyl operators might be used to construct \textit{mutually unbiased basis} \cite{klappenecker2005mutually,MUBs25,hiesmayr2021detecting}. 

\begin{definition}
Two orthonormal bases $B= \{ b_i\}_{i=1}^N$ and $C= \{ c_i\}_{i=1}^N$ in $N$-dimensional Hilbert space $\mathcal{H}_N$ are \textit{unbiased} iff the scalar product $|\langle b_i | c_i\rangle |^2 =1/N$ holds for any $b_i \in B,c_i \in C$. 
\end{definition}

There are not more than $N+1$ MUBs in dimension $N$ \cite{Ivonovic_1981,Wooters_1989}. 
For any prime dimension $N=p$, the eigenstates of the following $N+1$ operators
\begin{equation}
Z,\; X,\; XZ,\; XZ^2, \;\ldots,\; XZ^{N-1}
\label{lists}
\end{equation}
form a full set of $(N+1)$ MUBs, often called \textit{Heisenberg-Weyl MUBs}.

With a little effort, the above construction might be extended to any prime power dimension $N=p^n$. 
Unitary mulitilocal generalized Pauli operators 
$
X^{k_1}Z^{\ell_1} \otimes \cdots \otimes X^{k_n}Z^{\ell_n} ,
$ 
acting on a partitioned Hilbert space $\mathcal{H}^{\otimes n}_N$, can be partitioned into $N+1$ commuting classes of operators, in such a way that common eigenvectors of operators in each class form a basis, unbiased to any other one \cite{hiesmayr2021detecting}.  
Among all such bases exactly $p+1$ are \textit{local} bases, i.e. constituting vectors are separable with respect to the structure of a Hilbert space $\mathcal{H}^{\otimes n}_N$ \cite{LocalMubs}. 
In fact, the number of unbiased local bases cannot exceed $p+1$ \cite{Wooters_1989}.

Eigenvectors of Weyl-Heisenberg operators might be successfully used in order to construct SudoQ grids with maximum cardinalities and exceptional orthogonality relations.

\begin{proposition}
The following collection of vectors
\begin{equation}
\ket{v_{ijk\ell}} =  \ket{(j+k)_i}\otimes\ket{(i+\ell)_j},
\label{MUBLOC}
\end{equation}
in view of the notation introduced in Remark \ref{remark22},
forms a SudoQ grid of size $N^2$ with the maximal cardinality, where $\ket{j_k}$ are eigenvectors (\ref{vectt}) of the Heisenberg-Weyl operators. 
Moreover, if $N$ is prime, 
for each permutation $\sigma$ of $N$ indices, the following $N$ bases of $N^2$ dimensional space 
$\{\ket{v_{i\sigma (i) k\ell}}_{k,\ell =1}^N\}_{i=1}^N$ are mutually unbiased and local.  
\end{proposition}

\begin{proof}
Notice that by setting the eigenvectors $\ket{s_k}_{s=1}^N$ of $XZ^k$ operator in a column, we get a unitary $N\times N$ matrix. 
Therefore, by Proposition \ref{propp}, constructed grid is indeed a SudoQ. 
Moreover, two vectors $\ket{s_i}$ and $\ket{s'_{i'}}$ are either orthogonal, if $i=i'$, or unbiased, if $i\neq i'$. 
Since eigenvectors of $XZ^k$ for different $k$ are all different (see Appendix \ref{sec:WH_app}), $\ket{s_i} \neq \ket{s'_{i'}}$, except $s=s',i=i'$. 
Therefore, such a SudoQ grid is of the maximal cardinality. 
Notice that for a fixed values $i\neq i'$, vectors $\ket{p_i}$ and $\ket{p'_{i'}}$ are unbiased, i.e $\braket{p_i |p'_{i'}}=1/\sqrt{N}$. 
Therefore, for fixed values $i \neq i'$ and $j \neq j'$, the corresponding vectors are unbiased in a Hilbert space $\mathcal{H}_N\otimes \mathcal{H}_{N}$. 
\begin{align*}
\Big\vert & \braket{v_{ijk\ell} |v_{i'j'k'\ell'}} \Big\vert = \\
&\Big\vert\braket{(i+k)_i |(i'+k')_{i'} } \Big\vert\cdot
\Big\vert\braket{(j+\ell)_j |(j'+\ell')_{j'} } \Big\vert=
\frac{1}{N}.
\end{align*}

Therefore, for any permutation $\sigma$, the bases $\{\ket{v_{i\sigma (i) k\ell}}_{k,\ell =1}^N\}_{i=1}^N$ of $N^2$ dimensional space are mutually unbiased.    
Since all vectors in Eq.~(\ref{MUBLOC}) are of tensor product form, the corresponding bases are all local. 
\end{proof}

We present a simple construction of $9\times 9$ Heisenberg-Weyl SudoQ grid. 
Notice that it achieves the maximal cardinality, $c=81$, and the bases formed by vectors in appropriate blocks are unbiased.

\begin{figure*}[t]
\includegraphics[width=0.72\textwidth]{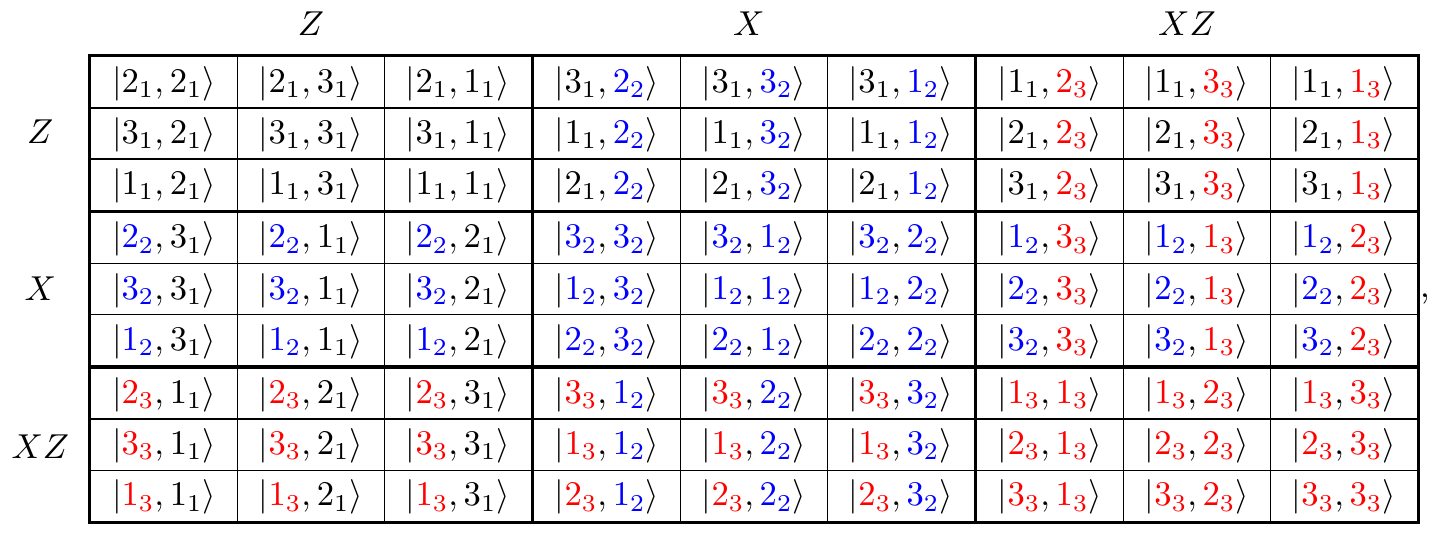}
\caption{The $9\times 9$ SudoQ grid with the maximal cardinality, $c=81$. 
Each entry is represented by the quantum state in a tensor product form $\ket{k_i} \otimes \ket{\ell_j} \in \mathcal{H}_3 \otimes \mathcal{H}_3$. 
Constituting vectors $\ket{k_i} , \ket{\ell_j}$ are $i$-th and $j$-th eigenvectors of Heisenberg-Weyl operators $XZ^{k-1}$ and $XZ^{\ell-1}$ respectively, see Eq. (\ref{vectt}) for their explicit form. Notice that all vectors in a given block are common eigenvectors of two operators indicated on the top and left side of a grid. 
Moreover, vectors from two blocks $(i,j)$ and $(i',j')$ are mutually unbiased for $i\neq i'$ and $j\neq j'$. }
\label{99}
\end{figure*}

\begin{example}{}
\upshape
For $N=3$, the Heisenberg-Weyl SudoQ construction leads to a $9\times 9$ solution of the maximal cardinality $81$ constructed with three triples of MUBs of size $9$ (in analogy to the two pairs of MUBs of order $4$ applied for the $4\times 4$ SudoQ from Ex. \ref{ex:4x4_card16}).

An $N=3$ dimensional operator $Z$ has the following eigenvectors:
\begin{equation}
\ket{1_1}=\ket{1}, \quad
\ket{2_1}=\ket{2}, \quad
\ket{3_1}=\ket{3} ,
\end{equation}
while eigenvectors of $ X,\: XZ$ are presented below, on the left and right side respectively:
{\footnotesize
\begin{alignat*}{4}
{\color{blue}\ket{1_2}}
&=\frac{1}{\sqrt{3}} (\ket{1}+\ket{2} +\ket{3}), 
\quad 
&{\color{red}\ket{1_3}}
&=\frac{1}{\sqrt{3}} (\ket{1}+\omega\ket{2} +\omega\ket{3}), 
\\
{\color{blue}\ket{2_2}}
&=\frac{1}{\sqrt{3}} (\ket{1}+ \omega \ket{2} + \omega^2\ket{3}), \quad 
&{\color{red}\ket{2_3}}
&=\frac{1}{\sqrt{3}} (\ket{1}+ \omega^2 \ket{2} + \ket{3}), 
\\
{\color{blue}\ket{3_2}}
&=\frac{1}{\sqrt{3}} (\ket{1}+ \omega^2 \ket{2} + \omega\ket{3}), \quad 
&{\color{red}\ket{3_3}}
&=\frac{1}{\sqrt{3}} (\ket{1}+ \ket{2} + \omega^2 \ket{3}).
\end{alignat*}
}
Note that the above eigenvectors of operators $ Z,\: X,\: XZ$ might be presented as columns of identity and the following two complex Hadamard matrices,
\[
\begin{bmatrix}
    1 & 0 & 0 \\
    0 & 1& 0 \\
    0 & 0 & 1
    \end{bmatrix},
    \quad
\frac{1}{\sqrt{3}}\begin{bmatrix}
    1 & 1 & 1 \\
    1 & \omega & \omega^2 \\
    1 & \omega^2 & \omega
    \end{bmatrix},
\quad 
\frac{1}{\sqrt{3}}\begin{bmatrix}
    1 & 1 & 1 \\
    \omega & \omega^2 & 1 \\
    \omega & 1 & \omega^2
    \end{bmatrix}.
\] 
The corresponding SudoQ grid is presented in Fig. (\ref{99}). 
\end{example}


\section{SudoQ cubes and hypercubes}\label{SudoqqCubes}

Remark \ref{remark22} presents exact orthogonality relations between $N^4$ vectors forming a SudoQ grid. 
We generalize the notion of SudoQ $2$-dimensional grid presented in Remark \ref{remark22} into higher dimensions, following already existing implementations of classical Sudoku cubes \cite{Lambert_2010}. 

\begin{definition}
\label{cube}
A \textit{SudoQ cube} is a collection of $N^6$ vectors 
$ \{ \ket{v}_{i_1 i_2 i_3  k_1 k_2 k_3} \} \in \mathcal{H}^{N^3}$, such that the marginal sets of $N^3$ vectors $ \{ \ket{v}_{i_1 i_2 i_3  k_1 k_2 k_3} \} \in \mathcal{H}^{N^3}$ constitute a basis for any fixed triplets of indices: $i_1, i_2 ,i_3$ or $k_1, i_2 ,i_3$ or $i_1, k_2 ,i_3$ or $i_1, i_2 ,k_3$.
\end{definition}

The simplest construction of such a SudoQ cube, comes from the classical construction of a Sudoku cube:
\begin{equation}
\ket{v}_{i_1 i_2 i_3  k_1 k_2 k_3}=
\ket{i_1 +k_1}\otimes \ket{i_2 +k_2}\otimes \ket{i_3 +k_3} \in \mathcal{H}_N^{\otimes 3},
\end{equation}
where addition is considered modulo $N$. 
One may easily check that conditions imposed in Definition \ref{cube} are satisfied. 
Notice, however, that such a SudoQ cube achieves the minimal possible cardinality $N^3$. 

We adapt the method of constructing SudoQ grids with a large cardinality to the $3$-dimensional setting. 

\begin{proposition}
\label{ppp}
Consider three families of $N\times N$ dimensional unitary operators: $\{U_i \}_{i=1}^N $, and $\{V_i \}_{i=1}^N $, and $\{W_i \}_{i=1}^N $. 
Denote by $\ket{u_{s}^{(i)}}$ the $s$-th column of $U_i$ operator ($\ket{v_{s}^{(i)}}$ from $V_i$ and $\ket{w_{s}^{(i)}}$ from $W_i$ respectively). 
The following collection of vectors
\begin{equation*}
\ket{x_{i_1 i_2 i_3  k_1 k_2 k_3}} := 
\ket{u_{k_1+ i_1}^{(i_2+i_3)}} \otimes 
\ket{v_{k_2+ i_2}^{(i_1+i_3)}} \otimes 
\ket{w_{k_3+ i_3}^{(i_1+i_2)}} 
\in \mathcal{H}_N^{\otimes 3},
\end{equation*}
constitute the SudoQ cube.
\end{proposition}

\begin{proof}
We shall verify that vectors 
$\ket{x_{i_1 i_2 i_3  k_1 k_2 k_3}}$ constitute a basis of a Hilbert space $\mathcal{H}_N^{\otimes 3}$ for four fixed triplets of indices specified in Definition \ref{cube}.

For three fixed parameters $i_1 ,i_2, i_3$, vectors 
$\ket{u_{k_1+ i_1}^{(i_2+i_3)}}$ are simply consecutive columns of a unitary operator $U_{i_2+i_3}$, and hence form a basis of $\mathcal{H}_N$. 
Similarly, vectors $\ket{v_{k_2+ i_2}^{(i_1+i_3)}}$ and $\ket{w_{k_3+ i_3}^{(i_1+i_2)}}$ form basis of $\mathcal{H}_N$, thus their tensor products 
form a basis of $\mathcal{H}_N^{\otimes 3}$. 

Observe that for fixed parameters $i_1 ,i_2, i_3, k_1$, the vectors 
\begin{equation}
\ket{u_{k_1+ i_1}^{(i_2+i_3)}} \otimes 
\ket{v_{k_2+ i_2}^{(i_1+i_3)}} \otimes 
\ket{w_{k_3+ i_3}^{(i_1+i_2)}},
\label{auxi}
\end{equation}
span the $N^2$-dimensional subspace, $\ket{u_{k_1+ i_1}^{(i_2+i_3)}} \otimes \mathcal{H}_N^{\otimes 2}$. 
Moreover, for fixed parameters $i_2, i_3, k_1$ and different values $i_1 =1,\ldots,N$, the vectors 
$\ket{u_{k_1+ i_1}^{(i_2+i_3)}} $ span the subspace $\mathcal{H}_N^{}$. 
Therefore, for fixed parameters $i_2, i_3, k_1$, corresponding vectors (\ref{auxi}) span $\mathcal{H}_N^{\otimes 3}$, and hence form its basis. 
Similar analysis might be performed with respect to fixed indices $i_1 ,k_2, i_3$ and $i_1 ,i_2, k_3$.
\end{proof}

Although provided construction is in analogy to the construction of a $2$-dimensional SudoQ grid in Proposition \ref{propp}, it does not achieve the maximal cardinality even for the appropriate choice of unitary matrices $\{U_i \}_{i=1}^N $, and $\{V_i \}_{i=1}^N $, and $\{W_i \}_{i=1}^N $ for all values of $N$. 
Indeed, the set of vectors $ \ket{v_{i_1 i_2 i_3  k_1 k_2 k_3}} $ constituting a SudoQ cube is in fact of the tensor product form:
\begin{align*}
\{&|x_{i_1 i_2 i_3  k_1 k_2 k_3} \rangle \}_{i_1, i_2, i_3 , k_1 ,k_2 ,k_3=1}^N =\\
&\quad\quad\{ u_{\ell_1}^{(j_1)} \otimes  
v_{\ell_2}^{(j_2)}  \otimes 
w_{\ell_3}^{(j_1+j_2 -2 j_3)} \}_{j_1, j_2, j_3 , \ell_1 ,\ell_2 ,\ell_3=1}^N ,
\end{align*}
which for even number $N$ contains exactly $N^6/2$ different vectors. 

In fact, the construction presented in Proposition \ref{ppp}, might be slightly modified:
\begin{equation*}
\ket{x_{i_1 i_2 i_3  k_1 k_2 k_3}} := 
\ket{u_{k_1+ i_1}^{(i_2+i_3)}} \otimes 
\ket{v_{k_2+ i_2}^{(i_1+i_3)}} \otimes 
\ket{w_{k_3+ i_3}^{(i_1+2i_2)}} 
\in \mathcal{H}_N^{\otimes 3},
\end{equation*}
for even values of $N$. 
The change in the upper index of the last product does not violate conditions imposed on the SudoQ cube. 
On the other hand, such a SudoQ cube achieves the maximal cardinality $N^6$ for three families of unitary matrices $\{U_i \}_{i=1}^N $, $\{V_i \}_{i=1}^N $ and $\{W_i \}_{i=1}^N $, such that in each family of matrices columns do not repeat (up to the multiplication by a phase factor).

Notice that Definition \ref{cube} of SudoQ cube might be generalized for an arbitrary $D$-dimensional hypercube. Furthermore, construction presented in Proposition \ref{ppp} can be generalized for construction of $D$-dimensional SudoQ hypercubes, with the maximal cardinality of $N^{2D}$. 

\section{Projective $t$-designs}
\label{comparison}

We shall investigate the problem of evenly spreading a set of unit vectors in a vector space. 
Let $X$ be a set of unit vectors in the Hilbert space $\mathcal{H}_d$. 
In general, vectors from any set $X$ satisfy the \textit{Welch bound} \cite{Welch}
\begin{equation}
W_t \coloneqq \dfrac{1}{|X|^2} \sum_{x,y \in X} |\langle x| y\rangle|^{2 t } 
\geq 
\dfrac{1}{\binom{d+t-1}{t}}
\label{welch}
\end{equation}
for any integer number $t \geq 0$. 
Observe that the following bound holds true
\begin{equation}
S_t:=\dfrac{1}{\binom{d+t-1}{t} W_t} \leq 1
\label{welch2}
\end{equation}
for any set of vectors. 
Sets of vectors which for a given integer $t$ saturate the bound above are called \textit{complex projective $t$-designs}. 
There is a remarkable feature of $t$-designs, namely the integration over the entire set of pure states, equivalent to the complex projection space $\mathbb{C}\mathbf{P}^{d-1} $, of any polynomial function $f$ of degree $t$ in state coefficients might be replaced by sampling over the states forming a $t$-designs. 
Indeed, for a $t$-design $X$ and any polynomial function $f\in \text{Hom} (t,t)$ of degree $t$ in both states and their conjugates the following equality holds,
\[
\frac{1}{\mu ( \mathbb{C}\mathbf{P}^{d-1} )}
\int_{\mathbb{C}\mathbf{P}^{d-1}}^{} 
f(x) d\mu (x)= 
\frac{1}{|X|}
\sum_{x\in X} f(x),
\]
where the integration is understood with respect to the Haar measure on a complex projective plane $\mathbb{C}\mathbf{P}^{d-1}$ \cite{klappenecker2005mutually}. 
For instance, any basis of $\mathbb{C}^{d}$ space forms a $1$-design. 
For a given set $X$ we define its \textit{angle set} by all possible scalar products, i.e. $\{ |\langle x| y\rangle|^{2 }_{i,j \in X} \}$. 

Complex projective $t$–designs are used in many branches of quantum information theory: quantum state
tomography, quantum fingerprinting, or quantum cryptography. 
In general, the larger $t$, the better given design approximates the state space. 
Except for the case $t=1,2$ the problem of constructing $t$-designs is not simple and no general construction is known. 

A complete set of $d+1$ MUBs, or $d^2$ SIC-POVMs in dimension $d$ comes with a rich combinatorial structure. 
Both are complex projective $2$-designs with an angle set $\{0,\frac{1}{d} \}$ for MUBs and $\{\frac{1}{d+1}\}$ for SIC-POVMs respectively. 

\begin{figure}
\includegraphics[width=0.44\textwidth]{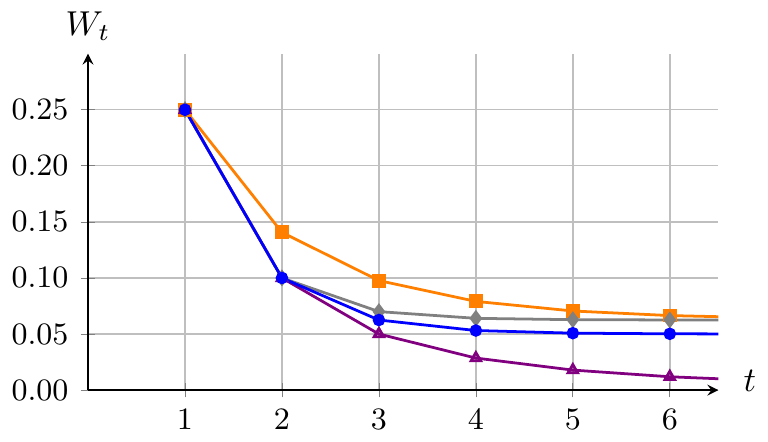}
\includegraphics[width=0.44\textwidth]{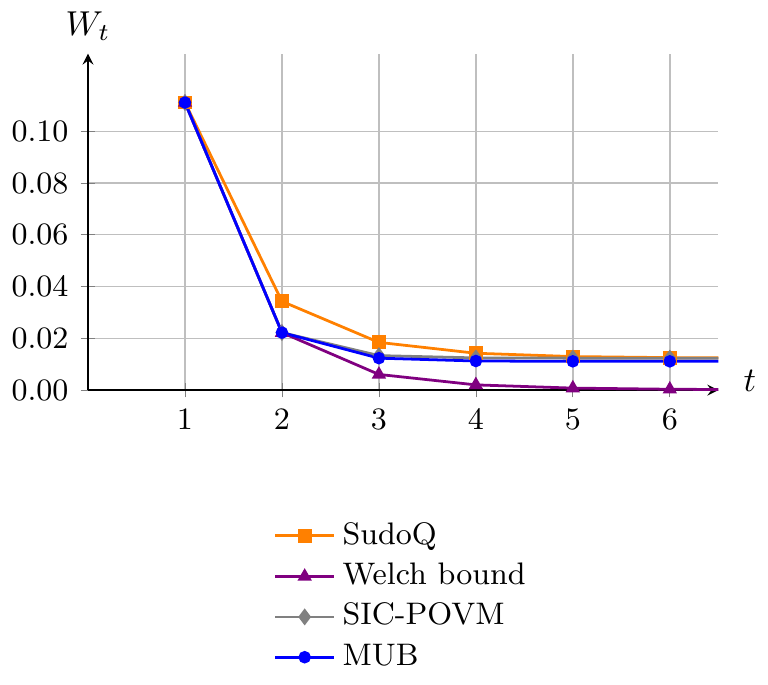}
\caption{Welch quantity $W_t$ for various projective designs as a function of the degree $t$ in Welch inequality for different projective designs: MUB, SIC-POVM, and SudoQ in dimensions $d=4$ on top and $d=9$ on bottom. 
MUBs and SIC-POVMs are $2$-designs, therefore the Welch bound is saturated for $t=1,2$. 
SudoQs are $1$-designs.}
\label{figComparison}
\end{figure}

In fact, presented Heisenberg-Weyl SudoQ exhibits a particular combinatorial structure. 
Notice that three orthogonality relations are possible, and the angle set is given by $\{0, \frac{1}{\sqrt{d}},\frac{1}{d}\}$, where $d=N^2$. 
Fig. \ref{figComparison} shows the value $W_t$ of the left-hand side of the Welch inequality (\ref{welch}) as a function of the degree $t$ of a design.

\begin{figure}
\includegraphics[width=0.45\textwidth]{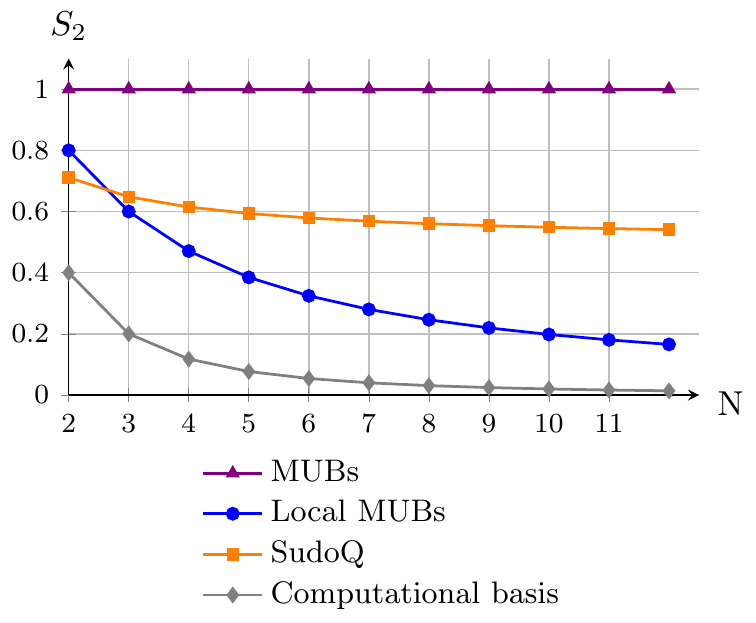}
\caption{Saturation $S_2$ of the Welch bound with $t=2$ as a function of the design parameter $N$ for different projective designs. Vectors from MUBs saturate the Welch bound with $t=2$. Among all local designs (SudoQ, local MUBs, and any distinguished basis), the Heisenberg-Weyl SudoQ design is closer to saturate the Welch inequality with $t=2$.
}
\label{figComparisonlocal}
\end{figure}

Table \ref{table} compares the Heisenberg-Weyl SudoQ against MUBs. 
Contrary to the MUBs design, the SudoQ design does not saturate Welch inequality for $t=2$. 
Nevertheless, in the Heisenberg-Weyl SudoQ design all vectors are separable. 
Considering only local vectors from both designs, the SudoQ design is closer to saturate Welch inequality with $t=2$ than the separable MUB related design, see Fig. \ref{figComparisonlocal}. 


\begin{table}
\centering 
\begin{tabular}{ | c ||  c|  c | } 
\hline
&  SudoQ &MUB \\ 
\hline
 Total number of states & $d^2$ &  $d(d+1)$ \\ 
\hline $\begin{gathered}\text{Maximal number of} \\ \text{separable states} \end{gathered}$
     & $d^2$ & $\sqrt{d}+1 $  \\
\hline
Angle set & $\{0, \frac{1}{\sqrt{d}},\frac{1}{d}\}$ &$\{0, \frac{1}{d}\}$  \\
\hline
$\begin{gathered} \text{Welch quantity } W_t \end{gathered}$ & $\dfrac{\big( \sqrt{d}^t+d -\sqrt{d} \big)^2}{d^{t+1}} $ & $\dfrac{1+d^{-t}}{d(1+d)}$ \\ 
\hline
\end{tabular}
\caption{Features of two projective designs formed by SudoQ vectors and MUBs vectors in squared dimension $d=N^2$ are compared, in particular saturation of the Welch bound.}
\label{table}
\end{table}

\section{Conclusions}
Sudoku is a very well-known classical game with interesting applications in science, i.e. in the study of dynamical systems and transient chaos~\cite{Sudoku_chaos}.
Following recent introduction of a quantum version of Sudoku \cite{Nechita2020SudoQA}, we have expanded this idea, defined the genuinely quantum SudoQ and analyzed its properties. More specifically, we found it useful to characterize any SudoQ solution by its cardinality, being the number of its vectors different up to the global phase. 
We have determined all admissible cardinalities, the parametrization of the solutions, and a proof of a conjecture from \cite{Nechita2020SudoQA} in the $4\times 4$ case.
In the general case of $N^2 \times N^2$ SudoQ we have found solutions of the maximal cardinality, $c=N^4$, together with genuinely quantum solutions of intermediate cardinalities.

The SudoQ problem has profound consequences for designing generalized quantum measurements, since each SudoQ pattern of order $N^2$ consisting of $N^4$ distinct vectors of size $N^2$ determines $3N^2$ orthogonal measurements, corresponding to $N^2$ rows, $N^2$ columns and $N^2$ blocks of the pattern.
Despite our general construction of the maximal cardinality, the full characterization of the admissible cardinalities in the case of an arbitrary dimension $N^2$ is still missing. 
The existence of SudoQ of different cardinalities provides an insight into bounds on a number of different measurements achievable from a given set of vectors.

In order to successfully construct SudoQ grids, we adapt the Heisenberg-Weyl method for constructing a complete set of $(d+1)$ MUBs. 
Our method is analogous to construction of MUBs in squared dimensions $d=N^2$. 
What is more, we have shown how to generalize this approach for SudoQ cubes and hypercubes. 

Similarly to MUBs, vectors from the Heisenberg-Weyl SudoQ grid form projective design. 
We compared both designs against the Welch bound, both of them saturating Welch inequality with $t=1$. 
In contrast to MUBs, vectors from SudoQ grid do not saturate Welch inequality for $t=2$, and hence form only $1$-design. 
Nevertheless, a SudoQ grid might be constructed in such a way that all related vectors are separable. 
Taking into account only local vectors from both designs, the SudoQ-related design is closer to meet the Welch inequality for $t=2$ than the separable MUB related design. 
Moreover, we show that SudoQ design outperforms any basis-related design. 
Therefore, the SudoQ-related design approximates the structure of Hilbert spaces best among all aforementioned local designs.

It is a pleasure to thank Ion Nechita for a fruitful correspondence. We acknowledge financial support by the National Science Center in Poland under the Maestro grant number DEC-2015/18/A/ST2/00274 and by the Foundation for Polish Science under the project Team-Net NTQC.

\appendix

\section{Proof of Theorem~\ref{cardinalities}}\label{sec:proof_cardinalities}
We present the proof of Theorem~\ref{cardinalities} concerning all possible cardinalities of $4\times 4$ SudoQ, which are $c = 4, 6, 8$ or $16$.
\begin{proof}
Without loss of generality we may consider grids with the upper left block composed of elements from the computational basis.
To fulfill the orthogonality relations in the first two rows and columns, the vectors in the upper right and lower left blocks must be of the form

\begin{equation}\label{generalgrid}
    \begin{tabular}{!{\vrule width 1pt}c|c!{\vrule width 1pt}c|c!{\vrule width 1pt}}
    \noalign{\hrule height 1pt}
    $\ket{1}$ & $\ket{2}$ & $\ket{a_{34}}$ & $\ket{b_{34}}$ \tabularnewline
    \hline
    $\ket{3}$ & $\ket{4}$ & $\ket{a_{12}}$ & $\ket{b_{12}}$ \tabularnewline
    \noalign{\hrule height 1pt}
    $\ket{a_{24}}$ & $\ket{a_{13}}$ & . & . \tabularnewline
    \hline
    $\ket{b_{24}}$ & $\ket{b_{13}}$ & . & . \tabularnewline
    \noalign{\hrule height 1pt}
    \end{tabular}\ .
\end{equation}

Let us start by noting that the above grid must contain even number of distinct vectors, since if we set any $\ket{a_{ij}}$ to be an element of the basis $\mathcal{B}$, then $\ket{b_{ij}}$ also must belong to $\mathcal{B}$ and vice versa. 
Consider all possible options of setting $\ket{a_{ij}}$ and $\ket{b_{ij}}$ pairwise to belong to $\mathcal{B}$.
Since there are 4 pairs, we have 5 distinct possibilities $\{0,1,2,3,4\}$ of pairs not in $\mathcal{B}$.

If we set all 4 pairs to be beyond $\mathcal{B}$, then the grid is either uniquely solvable or unsolvable. To illustrate this fact let us choose the set of vectors $\{\ket{a_{12}},\ket{a_{34}},\ket{a_{13}},\ket{a_{24}}\}$ and notice that we cannot write any of them as a linear combination of only two other. Therefore, the space spanned by this set is at least three dimensional. 
Since a vector orthogonal to the 3-dimensional subspace spanned vectors not in $\mathcal{B}$ cannot belong to $\mathcal{B}$, the only possibility not yet disproven is that all the missing vectors are non-computational and are superpositions of all $\{\ket{1},\ket{2},\ket{3},\ket{4}\}$, otherwise it would not be orthogonal to some of the already filled elements.
To sum up, the only case left are four new vectors, which form a base and thus the entire solution has cardinality $c=16$.

To disprove the possibility of 3 pairs outside of $\mathcal{B}$, let us choose without loosing generality that $\ket{a_{12}} = \ket{1}$.
Then the upper left element from the missing block must be a superposition of $\{\ket{2},\ket{3},\ket{4}\}$.
However, since $\ket{a_{13}} \notin \mathcal{B}$, the superposition narrows down to $\{\ket{2},\ket{4}\}$, and so it cannot be simultaneously orthogonal to $\ket{a_{24}}, \ket{a_{34}} \notin \mathcal{B}$.
Reasoning holds for all 3 pairs of vectors not from $\mathcal{B}$; thus, there are no solutions in the case of 3 pairs.


There are two distinct cases when 2 pairs of vectors $\notin \mathcal{B}$, either they are in the same block or in the different ones.
If we choose them from the different blocks then there exists an element in the missing block, which should be orthogonal to two different computational basis' vectors.
We can set it to be once again left upper element, with $\ket{a_{12}}=\ket{1}$ and $\ket{a_{13}}=\ket{3}$, as all the other combinations follow the same chain of reasoning.
Consequently, this element must be a superposition of $\{\ket{2},\ket{4}\}$, but then it cannot be orthogonal to $\ket{a_{24}}$ and $\ket{a_{34}}$ at the same time (since both are non-trivial superpositions), which shows that no solutions follow from this case.

If, on the other hand, two pairs of vectors not belonging to $\mathcal{B}$ are in the same block (and the vectors from the second one belong to $\mathcal{B}$), the grid might be solvable with $c=8$. To make it visible, let us set the vectors from upper right block $\ket{v_i} \notin \mathcal{B}$, and these from the lower left block to belong to $\mathcal{B}$. Without loss of generality we can assume that $\ket{a_{24}}=\ket{2}$ and $\ket{b_{24}}=\ket{4}$. Then choosing $\ket{a_{13}}=\ket{3}$ and $\ket{b_{13}}=\ket{1}$ one gets an unsolvable grid -- from orthogonality in rows it follows that the upper left element of the missing block must be a superposition of $\{\ket{1},\ket{4}\}$, but such a superposition cannot be orthogonal to $\ket{a_{12}}$ and $\ket{a_{34}}$ $\notin \mathcal{B}$. However, if we set $\ket{a_{13}}=\ket{1}$ and $\ket{b_{13}}=\ket{3}$, the considered element of the missing block must be a superposition of $\{\ket{3},\ket{4}\}$, and be orthogonal to both $\ket{a_{12}}$ and $\ket{a_{34}}$. There is only one normalized vector satisfying these conditions -- it is $\ket{b_{34}}$. Considering other elements of the missing block, we conclude that the grid is uniquely solvable, and the solution has 8 distinct vectors, and that the lower right block contains exactly the same vectors as the upper right one.


To check the possibility of a single pair of vectors beyond $\mathcal{B}$, consider the unsolvable grid from the last paragraph and assume that two of the vectors from the upper right block belong to $\mathcal{B}$. Then the grid is still unsolvable -- we can always find an entry of the missing block that cannot be filled. Nevertheless, if we make this change in the uniquely solvable grid, the grid is still uniquely solvable. The considerations similar to these from the last paragraph lead to the conclusion that $c=6$ in this case, and again, the vectors in the lower and upper right blocks are the same.

Finally, if we set all vectors from the upper right and lower left blocks to belong to $\mathcal{B}$, the solution has $c=4$. It can be concluded from the fact that choosing two or four vectors in one of these blocks to be non-computational implies the occurrence of the same vectors in the missing block. 
Therefore, vectors $\ket{v_i} \notin \mathcal{B}$ cannot occur in a single block, and in this situation vectors from the missing block must belong to $\mathcal{B}$.
\end{proof}

\section{Parametrization of $4\times 4$ SudoQ}\label{sec:16parametrization}
Let us recall the form of a general $4\times 4$ SudoQ solution given in Eq.~(\ref{generalsolution})
\begin{equation}\label{gensol}
    \begin{tabular}{!{\vrule width 1pt}c|c!{\vrule width 1pt}c|c!{\vrule width 1pt}}
    \noalign{\hrule height 1pt}
    $e_1$ & $e_2$ & $f_1$ & $f_2$ \tabularnewline
    \hline
    $e_3$ & $e_4$ & $f_3$ & $f_4$ \tabularnewline
    \noalign{\hrule height 1pt}
    $v_1$ & $v_2$ & $u_1$ & $u_2$ \tabularnewline
    \hline
    $v_3$ & $v_4$ & $u_3$ & $u_4$ \tabularnewline
    \noalign{\hrule height 1pt}
    \end{tabular}\ .
\end{equation}
We consider the grid with three blocks filled (Eq.~\ref{generalgrid})
\begin{equation}\label{gengrid}
    \begin{tabular}{!{\vrule width 1pt}c|c!{\vrule width 1pt}c|c!{\vrule width 1pt}}
    \noalign{\hrule height 1pt}
    $\ket{1}$ & $\ket{2}$ & $\ket{a_{34}}$ & $\ket{b_{34}}$ \tabularnewline
    \hline
    $\ket{3}$ & $\ket{4}$ & $\ket{a_{12}}$ & $\ket{b_{12}}$ \tabularnewline
    \noalign{\hrule height 1pt}
    $\ket{a_{24}}$ & $\ket{a_{13}}$ & . & . \tabularnewline
    \hline
    $\ket{b_{24}}$ & $\ket{b_{13}}$ & . & . \tabularnewline
    \noalign{\hrule height 1pt}
    \end{tabular}\ ,
\end{equation}
and analyze all the options of setting $\ket{a_{ij}}$ and $\ket{b_{ij}}$ to belong to the computational basis $\mathcal{B}$ that are allowed for in the proof of Thm.~\ref{cardinalities}. Comparing (\ref{gensol}) with (\ref{gengrid}) we conclude that the matrices $U_{ef}$ and $U_{ev}$ (\ref{matrixdef}) are of the following forms
\[
U_{ef}=
\begin{bmatrix}
0 & 0 & . & . \\
0 & 0  & . & . \\
.  & .  & 0  & 0 \\
. & .  & 0 & 0
\end{bmatrix},
\qquad
U_{ev}=
\begin{bmatrix}
0 & . & 0 & . \\
. & 0  & . & 0 \\
0  & .  & 0  & . \\
. & 0  & . & 0
\end{bmatrix},
\]
where dots represent non-zero entries.

First, let us consider the case when all of the vectors $\ket{a_{ij}}$ and $\ket{b_{ij}}$ from the upper right and lower left blocks of (\ref{gengrid}) do not belong to $\mathcal{B}$. Since $\ket{a_{ij}}$ and $\ket{b_{ij}}$ belong to the same subspace and are orthogonal to each other, they can be represented by antipodal points on the Bloch sphere. Therefore, we must search for the matrices $U_{ef}$ and $U_{ev}$ of the form
\[
\begin{split}
&U_{ef}=
\begin{bmatrix}
0 & 0 & \cos{\frac{\beta}{2}} & \sin{\frac{\beta}{2}} \\
0 & 0  & e^{i\phi}\sin{\frac{\beta}{2}} & -e^{i\phi}\cos{\frac{\beta}{2}} \\
\cos{\frac{\alpha}{2}}  & \sin{\frac{\alpha}{2}}  & 0  & 0 \\
e^{i\varphi}\sin{\frac{\alpha}{2}} & -e^{i\varphi}\cos{\frac{\alpha}{2}}  & 0 & 0
\end{bmatrix},\\
\\
&U_{ev}=
\begin{bmatrix}
0 & \cos{\frac{\delta}{2}} & 0 & \sin{\frac{\delta}{2}} \\
\cos{\frac{\gamma}{2}} & 0  & \sin{\frac{\gamma}{2}} & 0 \\
0  & e^{i\xi}\sin{\frac{\delta}{2}} & 0  & -e^{i\xi}\cos{\frac{\delta}{2}} \\
e^{i\eta}\sin{\frac{\gamma}{2}} & 0  & -e^{i\eta}\cos{\frac{\gamma}{2}} & 0
\end{bmatrix}.
\end{split}
\]
If the grid is solvable, then each set \{$f_1$, $f_3$, $v_1$, $v_2$\}, \{$f_2$, $f_4$, $v_1$, $v_2$\}, \{$f_1$, $f_3$, $v_3$, $v_4$\}, \{$f_1$, $f_3$, $v_1$, $v_2$\} from (\ref{gensol}) must span a space that is at most three dimensional. In our case it is equivalent to the condition that vectors in each set are linearly dependent. Let us break it down for the set:
\[
\begin{split}
    &f_1=
    \begin{bmatrix}
            0\\
            0\\
            \cos{\frac{\alpha}{2}}\\
            e^{i\varphi}\sin{\frac{\alpha}{2}}
    \end{bmatrix},
    \qquad
    f_3=
    \begin{bmatrix}
            \cos{\frac{\beta}{2}}\\
            e^{i\phi}\sin{\frac{\beta}{2}}\\
            0\\
            0
    \end{bmatrix},\\
    &v_1=
    \begin{bmatrix}
            0\\
            \cos{\frac{\gamma}{2}}\\
            0\\
            e^{i\eta}\sin{\frac{\gamma}{2}}
    \end{bmatrix},
    \qquad
    v_2=
    \begin{bmatrix}
            \cos{\frac{\delta}{2}}\\
            0\\
            e^{i\xi}\sin{\frac{\delta}{2}}\\
            0
    \end{bmatrix}.
\end{split}
\]
If these vectors are linearly dependent, there exist such $c_i$, $i$=1,2,3,4, that
\[
c_1f_1+c_2f_3+c_3v_1+c_4v_2=0
\]
holds, and at least one of the $c_i$ is non-zero. This equation leads to the condition
\[
\tan\frac{\alpha}{2}\tan\frac{\delta}{2}-e^{i(\phi+\eta-\varphi-\xi)}\tan\frac{\beta}{2}\tan\frac{\gamma}{2}=0.
\]
    Consideration of the other sets \{$f_i$, $f_j$, $v_k$, $v_l$\} implies the system of equations
\[
    \begin{cases}
        \tan\frac{\alpha}{2}\tan\frac{\delta}{2}-e^{i(\phi+\eta-\varphi-\xi)}\tan\frac{\beta}{2}\tan\frac{\gamma}{2}=0\\
        \cot\frac{\alpha}{2}\tan\frac{\delta}{2}-e^{i(\phi+\eta-\varphi-\xi)}\cot\frac{\beta}{2}\tan\frac{\gamma}{2}=0\\
        \tan\frac{\alpha}{2}\cot\frac{\delta}{2}-e^{i(\phi+\eta-\varphi-\xi)}\tan\frac{\beta}{2}\cot\frac{\gamma}{2}=0\\
        \cot\frac{\alpha}{2}\cot\frac{\delta}{2}-e^{i(\phi+\eta-\varphi-\xi)}\cot\frac{\beta}{2}\cot\frac{\gamma}{2}=0
    \end{cases},
\]
with the solution 
\[
e^{i(\phi+\eta-\varphi-\xi)}=1, \qquad \alpha=\beta, \qquad \gamma=\delta.
\]
    Therefore, we can rewrite matrices $U_{ef}$ and $U_{ev}$ in the form
\newline
\newline
\begin{widetext}
\[
\begin{split}
&U_{ef}=
\begin{bmatrix}
0 & 0 & \cos{\frac{\alpha}{2}} & \sin{\frac{\alpha}{2}} \\
0 & 0  & e^{i\phi}\sin{\frac{\alpha}{2}} & -e^{i\phi}\cos{\frac{\alpha}{2}} \\
\cos{\frac{\alpha}{2}}  & \sin{\frac{\alpha}{2}}  & 0  & 0 \\
e^{i\varphi}\sin{\frac{\alpha}{2}} & -e^{i\varphi}\cos{\frac{\alpha}{2}}  & 0 & 0
\end{bmatrix},\\
\\
&U_{ev}=
\begin{bmatrix}
0 & \cos{\frac{\gamma}{2}} & 0 & \sin{\frac{\gamma}{2}} \\
\cos{\frac{\gamma}{2}} & 0  & \sin{\frac{\gamma}{2}} & 0 \\
0  & e^{i(\phi+\eta-\varphi)}\sin{\frac{\gamma}{2}} & 0  & -e^{i(\phi+\eta-\varphi)}\cos{\frac{\gamma}{2}} \\
e^{i\eta}\sin{\frac{\gamma}{2}} & 0  & -e^{i\eta}\cos{\frac{\gamma}{2}} & 0
\end{bmatrix}.
\end{split}
\]
\end{widetext}

Since appropriate sets of vectors from the upper right and lower left block span a three dimensional space, the grid is either uniquely solvable or unsolvable. Choosing the following matrix $U_{eu}$ we obtain a proper solution
\begin{widetext}
\[
U_{eu}=
    \begin{bmatrix}                      \sin\frac{\alpha}{2}\sin\frac{\gamma}{2} &         \cos\frac{\alpha}{2}\sin\frac{\gamma}{2} &
    \sin\frac{\alpha}{2}\cos\frac{\gamma}{2} &
    \cos\frac{\alpha}{2}\cos\frac{\gamma}{2}\\
    -e^{i\phi}\cos\frac{\alpha}{2}\sin\frac{\gamma}{2} &
    e^{i\phi}\sin\frac{\alpha}{2}\sin\frac{\gamma}{2} &
    -e^{i\phi}\cos\frac{\alpha}{2}\cos\frac{\gamma}{2} &
    e^{i\phi}\sin\frac{\alpha}{2}\cos\frac{\gamma}{2}\\
    -e^{i(\phi+\eta-\varphi)}\sin\frac{\alpha}{2}\cos\frac{\gamma}{2} &
    -e^{i(\phi+\eta-\varphi)}\cos\frac{\alpha}{2}\cos\frac{\gamma}{2} &
    e^{i(\phi+\eta-\varphi)}\sin\frac{\alpha}{2}\sin\frac{\gamma}{2} &
    e^{i(\phi+\eta-\varphi)}\cos\frac{\alpha}{2}\sin\frac{\gamma}{2}\\
    e^{i(\phi+\eta)}\cos\frac{\alpha}{2}\cos\frac{\gamma}{2} &
    -e^{i(\phi+\eta)}\sin\frac{\alpha}{2}\cos\frac{\gamma}{2} &
    -e^{i(\phi+\eta)}\cos\frac{\alpha}{2}\sin\frac{\gamma}{2} &
    e^{i(\phi+\eta)}\sin\frac{\alpha}{2}\sin\frac{\gamma}{2}
    \end{bmatrix}.
\]
\end{widetext}
Since for every SudoQ with $c=16$ all of the vectors $\ket{a_{ij}}$ and $\ket{b_{ij}}$ from the corresponding grid (\ref{gengrid}) do not belong to $\mathcal{B}$, this is the full parametrization of the above case. Note that if we set either $\alpha$ or $\gamma$ to be $0$ or $\pi$, we obtain a solution with $c=8$. If we set both of them to be such, we get a solution with $c=4$.
\newline
\newline
Now let us consider another case allowed by Theorem~\ref{cardinalities}, when the vectors from the lower left block of (\ref{gengrid}) belong to $\mathcal{B}$ while the vectors from the upper right block do not.
In this situation the grid (\ref{gengrid}) transforms to one of the following grids
\begin{equation}
\begin{split}
    &\begin{tabular}{!{\vrule width 1pt}c|c!{\vrule width 1pt}c|c!{\vrule width 1pt}}
    \noalign{\hrule height 1pt}
    $\ket{1}$ & $\ket{2}$ & $\ket{a_{34}}$ & $\ket{b_{34}}$ \tabularnewline
    \hline
    $\ket{3}$ & $\ket{4}$ & $\ket{a_{12}}$ & $\ket{b_{12}}$ \tabularnewline
    \noalign{\hrule height 1pt}
    $\ket{2}$ & $\ket{1}$ & . & . \tabularnewline
    \hline
    $\ket{4}$ & $\ket{3}$ & . & . \tabularnewline
    \noalign{\hrule height 1pt}
    \end{tabular}\ ,\qquad
    \begin{tabular}{!{\vrule width 1pt}c|c!{\vrule width 1pt}c|c!{\vrule width 1pt}}
    \noalign{\hrule height 1pt}
    $\ket{1}$ & $\ket{2}$ & $\ket{a_{34}}$ & $\ket{b_{34}}$ \tabularnewline
    \hline
    $\ket{3}$ & $\ket{4}$ & $\ket{a_{12}}$ & $\ket{b_{12}}$ \tabularnewline
    \noalign{\hrule height 1pt}
    $\ket{4}$ & $\ket{3}$ & . & . \tabularnewline
    \hline
    $\ket{2}$ & $\ket{1}$ & . & . \tabularnewline
    \noalign{\hrule height 1pt}
    \end{tabular}\ ,\\ \\
    &\begin{tabular}{!{\vrule width 1pt}c|c!{\vrule width 1pt}c|c!{\vrule width 1pt}}
    \noalign{\hrule height 1pt}
    $\ket{1}$ & $\ket{2}$ & $\ket{a_{34}}$ & $\ket{b_{34}}$ \tabularnewline
    \hline
    $\ket{3}$ & $\ket{4}$ & $\ket{a_{12}}$ & $\ket{b_{12}}$ \tabularnewline
    \noalign{\hrule height 1pt}
    $\ket{2}$ & $\ket{3}$ & . & . \tabularnewline
    \hline
    $\ket{4}$ & $\ket{1}$ & . & . \tabularnewline
    \noalign{\hrule height 1pt}
    \end{tabular}\ ,\qquad
    \begin{tabular}{!{\vrule width 1pt}c|c!{\vrule width 1pt}c|c!{\vrule width 1pt}}
    \noalign{\hrule height 1pt}
    $\ket{1}$ & $\ket{2}$ & $\ket{a_{34}}$ & $\ket{b_{34}}$ \tabularnewline
    \hline
    $\ket{3}$ & $\ket{4}$ & $\ket{a_{12}}$ & $\ket{b_{12}}$ \tabularnewline
    \noalign{\hrule height 1pt}
    $\ket{4}$ & $\ket{1}$ & . & . \tabularnewline
    \hline
    $\ket{2}$ & $\ket{3}$ & . & . \tabularnewline
    \noalign{\hrule height 1pt}
    \end{tabular}\ .
\end{split}
\end{equation}

However, observe that the lower two of the above grids are unsolvable -- it is sufficient to check that the set $\{\ket{2},\ket{3},\ket{a_{34}},\ket{a_{12}}\}$ span a four dimensional space, and therefore there is at least one entry in the lower right block that cannot be filled.

On the other hand, the first two grids are either uniquely solvable or unsolvable, since each of the sets $\{\ket{2},\ket{1},\ket{a_{34}},\ket{a_{12}}\}$, $\{\ket{2},\ket{1},\ket{b_{34}},\ket{b_{12}}\}$, $\{\ket{4},\ket{3},\ket{a_{34}},\ket{a_{12}}\}$ and $\{\ket{4},\ket{3},\ket{b_{34}},\ket{b_{12}}\}$ spans a three dimensional space. 
To show that these grids are solvable, it is enough to write down their solutions, which are respectively
\begin{equation}\label{card8sol}
\begin{tabular}{!{\vrule width 1pt}c|c!{\vrule width 1pt}c|c!{\vrule width 1pt}}
    \noalign{\hrule height 1pt}
    $\ket{1}$ & $\ket{2}$ & $\ket{a_{34}}$ & $\ket{b_{34}}$ \tabularnewline
    \hline
    $\ket{3}$ & $\ket{4}$ & $\ket{a_{12}}$ & $\ket{b_{12}}$ \tabularnewline
    \noalign{\hrule height 1pt}
    $\ket{2}$ & $\ket{1}$ & $\ket{b_{34}}$ & $\ket{a_{34}}$ \tabularnewline
    \hline
    $\ket{4}$ & $\ket{3}$ & $\ket{b_{12}}$ & $\ket{a_{12}}$ \tabularnewline
    \noalign{\hrule height 1pt}
    \end{tabular}\ ,\qquad
    \begin{tabular}{!{\vrule width 1pt}c|c!{\vrule width 1pt}c|c!{\vrule width 1pt}}
    \noalign{\hrule height 1pt}
    $\ket{1}$ & $\ket{2}$ & $\ket{a_{34}}$ & $\ket{b_{34}}$ \tabularnewline
    \hline
    $\ket{3}$ & $\ket{4}$ & $\ket{a_{12}}$ & $\ket{b_{12}}$ \tabularnewline
    \noalign{\hrule height 1pt}
    $\ket{4}$ & $\ket{3}$ & $\ket{b_{12}}$ & $\ket{a_{12}}$ \tabularnewline
    \hline
    $\ket{2}$ & $\ket{1}$ & $\ket{a_{34}}$ & $\ket{b_{34}}$ \tabularnewline
    \noalign{\hrule height 1pt}
    \end{tabular}\ .
\end{equation}
\mbox\newline
Since $\ket{a_{ij}}$ and $\ket{b_{ij}}$ are represented by antipodal points on the Bloch sphere, the matrices $U_{ef}$, $U_{ev}$ and $U_{eu}$ must be of the form
\begin{equation}\label{card8:param1}
\begin{split}
&U_{ef}=
\begin{bmatrix}
0 & 0 & \cos{\frac{\beta}{2}} & \sin{\frac{\beta}{2}} \\
0 & 0  & e^{i\phi}\sin{\frac{\beta}{2}} & -e^{i\phi}\cos{\frac{\beta}{2}} \\
\cos{\frac{\alpha}{2}}  & \sin{\frac{\alpha}{2}}  & 0  & 0 \\
e^{i\varphi}\sin{\frac{\alpha}{2}} & -e^{i\varphi}\cos{\frac{\alpha}{2}}  & 0 & 0
\end{bmatrix},
\\
\\
&U_{ev}=
\begin{bmatrix}
0 & 1 & 0 & 0 \\
1 & 0  & 0 & 0 \\
0  & 0 & 0  & 1 \\
0 & 0  & 1 & 0
\end{bmatrix},
\\
\\
&U_{eu}=
\begin{bmatrix}
0 & 0 & \sin{\frac{\beta}{2}} & \cos{\frac{\beta}{2}} \\
0 & 0 & -e^{i\phi}\cos{\frac{\beta}{2}} & e^{i\phi}\sin{\frac{\beta}{2}} \\
\sin{\frac{\alpha}{2}}  & \cos{\frac{\alpha}{2}}  & 0  & 0 \\
-e^{i\varphi}\cos{\frac{\alpha}{2}}  & e^{i\varphi}\sin{\frac{\alpha}{2}} & 0 & 0
\end{bmatrix}
\end{split}
\end{equation}
in the first case, and
\begin{equation}\label{card8:param2}
\begin{split}
&U_{ef}=
\begin{bmatrix}
0 & 0 & \cos{\frac{\beta}{2}} & \sin{\frac{\beta}{2}} \\
0 & 0  & e^{i\phi}\sin{\frac{\beta}{2}} & -e^{i\phi}\cos{\frac{\beta}{2}} \\
\cos{\frac{\alpha}{2}}  & \sin{\frac{\alpha}{2}}  & 0  & 0 \\
e^{i\varphi}\sin{\frac{\alpha}{2}} & -e^{i\varphi}\cos{\frac{\alpha}{2}}  & 0 & 0
\end{bmatrix},
\\
\\
&U_{ev}=
\begin{bmatrix}
0 & 0 & 0 & 1 \\
0 & 0  & 1 & 0 \\
0  & 1 & 0  & 0 \\
1 & 0  & 0 & 0
\end{bmatrix},
\\
\\
&U_{eu}=
\begin{bmatrix}
\sin{\frac{\beta}{2}} & \cos{\frac{\beta}{2}} & 0 & 0  \\
-e^{i\phi}\cos{\frac{\beta}{2}} & e^{i\phi}\sin{\frac{\beta}{2}} & 0 & 0  \\
  0  & 0 & \sin{\frac{\alpha}{2}}  & \cos{\frac{\alpha}{2}}\\
0 & 0 & -e^{i\varphi}\cos{\frac{\alpha}{2}}  & e^{i\varphi}\sin{\frac{\alpha}{2}}
\end{bmatrix}
\end{split}
\end{equation}
in the second one. 

The situation, in which the vectors from the lower left block of (\ref{gengrid}) are non-computational, and the upper right block contains only the elements of the computational basis, is analogous to the one examined above. In this case, the two possible solutions are
\begin{equation}
    \begin{tabular}{!{\vrule width 1pt}c|c!{\vrule width 1pt}c|c!{\vrule width 1pt}}
    \noalign{\hrule height 1pt}
    $\ket{1}$ & $\ket{2}$ & $\ket{3}$ & $\ket{4}$ \tabularnewline
    \hline
    $\ket{3}$ & $\ket{4}$ & $\ket{1}$ & $\ket{2}$ \tabularnewline
    \noalign{\hrule height 1pt}
    $\ket{a_{24}}$ & $\ket{a_{13}}$ & $\ket{b_{24}}$ & $\ket{b_{13}}$ \tabularnewline
    \hline
    $\ket{b_{24}}$ & $\ket{b_{13}}$ & $\ket{a_{24}}$ & $\ket{a_{13}}$ \tabularnewline
    \noalign{\hrule height 1pt}
    \end{tabular}\ , \qquad
    \begin{tabular}{!{\vrule width 1pt}c|c!{\vrule width 1pt}c|c!{\vrule width 1pt}}
    \noalign{\hrule height 1pt}
    $\ket{1}$ & $\ket{2}$ & $\ket{4}$ & $\ket{3}$ \tabularnewline
    \hline
    $\ket{3}$ & $\ket{4}$ & $\ket{2}$ & $\ket{1}$ \tabularnewline
    \noalign{\hrule height 1pt}
    $\ket{a_{24}}$ & $\ket{a_{13}}$ & $\ket{b_{13}}$ & $\ket{b_{24}}$ \tabularnewline
    \hline
    $\ket{b_{24}}$ & $\ket{b_{13}}$ & $\ket{a_{13}}$ & $\ket{a_{24}}$ \tabularnewline
    \noalign{\hrule height 1pt}
    \end{tabular}\ .
\end{equation}
The corresponding matrices $U_{ef}$, $U_{ev}$ and $U_{eu}$ are respectively
\begin{equation}\label{card8:param3}
\begin{split}
&U_{ef}=
\begin{bmatrix}
0 & 0 & 1 & 0 \\
0 & 0  & 0 & 1 \\
1  & 0 & 0  & 0 \\
0 & 1  & 0 & 0
\end{bmatrix},
\\
\\
&U_{ev}=
\begin{bmatrix}
0 & \cos{\frac{\beta}{2}} & 0 & \sin{\frac{\beta}{2}} \\
\cos{\frac{\alpha}{2}} & 0  & \sin{\frac{\alpha}{2}} & 0 \\
0  & e^{i\phi}\sin{\frac{\beta}{2}} & 0  & -e^{i\phi}\cos{\frac{\beta}{2}} \\
e^{i\varphi}\sin{\frac{\alpha}{2}} & 0  & -e^{i\varphi}\cos{\frac{\alpha}{2}} & 0
\end{bmatrix},
\\
\\
&U_{eu}=
\begin{bmatrix}
0 & \sin{\frac{\beta}{2}} & 0 & \cos{\frac{\beta}{2}} \\
\sin{\frac{\alpha}{2}} & 0  & \cos{\frac{\alpha}{2}} & 0 \\
0  & -e^{i\phi}\cos{\frac{\beta}{2}} & 0  & e^{i\phi}\sin{\frac{\beta}{2}} \\
-e^{i\varphi}\cos{\frac{\alpha}{2}} & 0  & e^{i\varphi}\sin{\frac{\alpha}{2}} & 0
\end{bmatrix},
\end{split}
\end{equation}
and
\begin{equation}\label{card8:param4}
\begin{split}
&U_{ef}=
\begin{bmatrix}
0 & 0 & 0 & 1 \\
0 & 0  & 1 & 0 \\
0  & 1 & 0  & 0 \\
1 & 0  & 0 & 0
\end{bmatrix},
\\
\\
&U_{ev}=
\begin{bmatrix}
0 & \cos{\frac{\beta}{2}} & 0 & \sin{\frac{\beta}{2}} \\
\cos{\frac{\alpha}{2}} & 0  & \sin{\frac{\alpha}{2}} & 0 \\
0  & e^{i\phi}\sin{\frac{\beta}{2}} & 0  & -e^{i\phi}\cos{\frac{\beta}{2}} \\
e^{i\varphi}\sin{\frac{\alpha}{2}} & 0  & -e^{i\varphi}\cos{\frac{\alpha}{2}} & 0
\end{bmatrix},
\\
\\
&U_{eu}=
\begin{bmatrix}
\sin{\frac{\beta}{2}} & 0 & \cos{\frac{\beta}{2}} & 0 \\
0  & \sin{\frac{\alpha}{2}} & 0 & \cos{\frac{\alpha}{2}} \\
-e^{i\phi}\cos{\frac{\beta}{2}} & 0 & e^{i\phi}\sin{\frac{\beta}{2}} & 0  \\
0  & -e^{i\varphi}\cos{\frac{\alpha}{2}} & 0 & e^{i\varphi}\sin{\frac{\alpha}{2}}
\end{bmatrix}.
\end{split}
\end{equation}

Setting $\alpha$ or $\beta$ equal to $0$ or $\pi$ (points on the poles of the Bloch sphere) allows us to obtain solutions with $c=6$. Notice that by this procedure we can attain every solution whose corresponding grid (\ref{gengrid}) has exactly one pair of non-computational vectors. Moreover, analyzing the form of the solutions (\ref{card8sol}) we conclude that it is impossible to have vectors outside $\mathcal{B}$ in a single block of the SudoQ. Therefore, if all the vectors $\ket{a_{ij}}$ and $\ket{b_{ij}}$ from the grid (\ref{gengrid}) are computational, the vectors from the lower right block must also be computational.
\newline
\newline
To sum up, we analyzed all the configurations of the grid (\ref{gengrid}), and parametrized its genuinely quantum solutions ($c>4$). By the action of a unitary matrix on all of the vectors it is possible to obtain all genuinely quantum solutions of a $4\times 4$ SudoQ.

\section{Shannon Entropy of SudoQ designs}\label{sec:shannon}
\begin{definition}
The Shannon entropy of a normalized vector $\ket{\psi}$ represented in the computational basis $\mathcal{B} = \{\ket{i}\}$ reads
\[
S(\ket{\psi})=-\sum_i\abs{\braket{i|\psi}}^2\log_2\left(\abs{\braket{i|\psi}}^2\right).
\]
The entropy of a SudoQ is defined as the sum of entropies of all normalized vectors forming the pattern.
\end{definition}

The Shannon entropy of the $4\times 4$ solution of cardinality 16 parametrized in Appendix \ref{sec:16parametrization} equals
\[
S(p,q)=2h_2(p)+2h_2(q)+4h_4(p,q),
\]
where $p=\cos^2\left(\frac{\alpha}{2}\right)$, $q=\cos^2\left(\frac{\gamma}{2}\right)$, $h_2(x)$ is the binary entropy, and 
\[
\begin{split}
    h_4(x,y)=&-xy\log_2(xy)-x(1-y)\log_2(x(1-y))\\&-(1-x)y\log_2((1-x)y)\\&-(1-x)(1-y)\log_2((1-x)(1-y)).
\end{split}
\]
This entropy has its maximum at $p=0.5$ and $q=0.5$. Therefore, Ex.~\ref{ex:4x4_card16}, which is in relation to two pairs of unbiased bases of size 4, is the one admitting the maximal entropy.

\section{Proof of Theorem~\ref{thm:4_clues}}\label{sec:proof_thm}
In the following appendix we shall prove Theorem~\ref{thm:4_clues}, which states:

\emph{For any 4$\times$4 quantum square with $c=16$, it is possible to construct a corresponding uniquely solvable grid with 4 clues.}
\begin{proof}
Consider the general form of the Sudoku solution
\begin{equation}
    \begin{tabular}{!{\vrule width 1pt}c|c!{\vrule width 1pt}c|c!{\vrule width 1pt}}
    \noalign{\hrule height 1pt}
    $e_1$ & $e_2$ & $f_1$ & $f_2$ \tabularnewline
    \hline
    $e_3$ & $e_4$ & $f_3$ & $f_4$ \tabularnewline
    \noalign{\hrule height 1pt}
    $v_1$ & $v_2$ & $u_1$ & $u_2$ \tabularnewline
    \hline
    $v_3$ & $v_4$ & $u_3$ & $u_4$ \tabularnewline
    \noalign{\hrule height 1pt}
    \end{tabular}\ .
\end{equation}
We will prove that the following grid is uniquely solvable
\begin{equation}
    \begin{tabular}{!{\vrule width 1pt}c|c!{\vrule width 1pt}c|c!{\vrule width 1pt}}
    \noalign{\hrule height 1pt}
    $e_1$ &   &   &   \tabularnewline
    \hline
      &   & $f_3$ &   \tabularnewline
    \noalign{\hrule height 1pt}
      & $v_2$ &   &   \tabularnewline
    \hline
      &   &   & $u_4$ \tabularnewline
    \noalign{\hrule height 1pt}
    \end{tabular}\ .
\end{equation}
\newline
\newline
First we conclude from the parametrization included in Appendix \ref{sec:16parametrization} that $e_1$, $f_3$ and $u_4$ are linearly independent, and therefore span a 3-dimensional subspace (the same holds true for $e_1$, $v_2$ and $u_4$). Hence, we can fill the upper right corner of the grid uniquely
\begin{equation}
    \begin{tabular}{!{\vrule width 1pt}c|c!{\vrule width 1pt}c|c!{\vrule width 1pt}}
    \noalign{\hrule height 1pt}
    $e_1$ &   &   &  $f_2$  \tabularnewline
    \hline
      &   & $f_3$ &   \tabularnewline
    \noalign{\hrule height 1pt}
      & $v_2$ &   &   \tabularnewline
    \hline
      &   &   & $u_4$ \tabularnewline
    \noalign{\hrule height 1pt}
    \end{tabular}\ ,
\end{equation}
where $f_2$ is orthogonal to the subspace spanned by $e_1$, $f_3$ and $u_4$. Since $f_2$ is orthogonal to $e_1$ and $f_3$, these three vectors also span a 3-dimensional subspace, which leads to the unique $f_1$. The same holds true for $f_4$
\begin{equation}
    \begin{tabular}{!{\vrule width 1pt}c|c!{\vrule width 1pt}c|c!{\vrule width 1pt}}
    \noalign{\hrule height 1pt}
    $e_1$ &   & $f_1$ &  $f_2$  \tabularnewline
    \hline
      &   & $f_3$ & $f_4$  \tabularnewline
    \noalign{\hrule height 1pt}
      & $v_2$ &   &   \tabularnewline
    \hline
      &   &   & $u_4$ \tabularnewline
    \noalign{\hrule height 1pt}
    \end{tabular}\ .
\end{equation}

Consequently, $e_2$ and $f_4$ are unique
\begin{equation}
    \begin{tabular}{!{\vrule width 1pt}c|c!{\vrule width 1pt}c|c!{\vrule width 1pt}}
    \noalign{\hrule height 1pt}
    $e_1$ & $e_2$  & $f_1$ &  $f_2$  \tabularnewline
    \hline
      &   & $f_3$ & $f_4$  \tabularnewline
    \noalign{\hrule height 1pt}
      & $v_2$ &   & $u_2$  \tabularnewline
    \hline
      &   &   & $u_4$ \tabularnewline
    \noalign{\hrule height 1pt}
    \end{tabular}\ .
\end{equation}
By an analogous chain of reasoning we can fill in all the elements below diagonal
\[
    \begin{tabular}{!{\vrule width 1pt}c|c!{\vrule width 1pt}c|c!{\vrule width 1pt}}
    \noalign{\hrule height 1pt}
    $e_1$ & $e_2$  & $f_1$ &  $f_2$  \tabularnewline
    \hline
    $e_3$  &   & $f_3$ & $f_4$  \tabularnewline
    \noalign{\hrule height 1pt}
    $v_1$  & $v_2$ &   & $u_2$  \tabularnewline
    \hline
    $v_3$  & $v_4$  & $u_3$  & $u_4$ \tabularnewline
    \noalign{\hrule height 1pt}
    \end{tabular}\ .
\]
Obviously, the last two entries are also given uniquely
\[
    \begin{tabular}{!{\vrule width 1pt}c|c!{\vrule width 1pt}c|c!{\vrule width 1pt}}
    \noalign{\hrule height 1pt}
    $e_1$ & $e_2$  & $f_1$ &  $f_2$  \tabularnewline
    \hline
    $e_3$  & $e_4$  & $f_3$ & $f_4$  \tabularnewline
    \noalign{\hrule height 1pt}
    $v_1$  & $v_2$ & $u_1$  & $u_2$  \tabularnewline
    \hline
    $v_3$  & $v_4$  & $u_3$  & $u_4$ \tabularnewline
    \noalign{\hrule height 1pt}
    \end{tabular}\ .
\]
\end{proof}

\section{Proof of Proposition \ref{Nechita_conjecture}}\label{sec:conjecture_Nechita}

Suppose that a classical $4\times 4$ Sudoku grid has a unique classical solution.
By associating vectors $i \mapsto \ket{i}$ to the clues in an initial Sudoku grid, we obtain a SudoQ grid. 
We shall show that it is uniquely solvable by proving that no genuinely quantum solutions exist. 
According to Theorem \ref{cardinalities}, any SudoQ solution of size $2^2 = 4$ has one of the cardinalities $c=4,6,8,16$. 

At first, let us assume that the solution is of cardinality $c=16$. Notice that by applying a unitary operation to all the vectors from the solution parametrized by us (\ref{parametrisation}), we can obtain any solution of $c=16$. Since unitary operations do not change the scalar product, and each two clues from the original grid has the scalar product either 0 or 1, we conclude that all the clues must belong to the same row, column or block (indeed, using (\ref{parametrisation}) we can check that the scalar product of vectors which do not belong to the same row, column or block, is different than 0 and 1). However, the classical grid with all the clues in the same row, column or block is never uniquely solvable, which contradicts our assumption. Therefore, the solution cannot be of the cardinality $c=16$.


Secondly, cardinality $c=8$ is also excluded. Reasoning analogous to argument from the previous paragraph (based on the parametrization (\ref{card8:param1},\ref{card8:param2},\ref{card8:param3},\ref{card8:param4}) of solutions of cardinality $c=8$), leads to the conclusion that all the clues must belong to the same two rows or columns. Again, the grid with two rows/columns filled is not uniquely solvable, since we can always transpose the remaining two rows/columns obtaining two distinct solutions.


What is more, the cardinality $c=6$ might also be excluded. In this case the clues in the original grid can be placed everywhere, except four entries. However, even if we fill all the other entries (in accordance with the parametrization of $c=6$ solutions), all the remaining four vectors will belong to the same two dimensional subspace spanned by two computational vectors. Thus, they can be chosen to be computational vectors in two different ways, from which we conclude that the original grid was not uniquely solvable. To make it more prominent, let us choose the parametrization (\ref{card8:param1}) and set $\beta=0$. The corresponding solution is
\begin{equation}
    \begin{tabular}{!{\vrule width 1pt}c|c!{\vrule width 1pt}c|c!{\vrule width 1pt}}
    \noalign{\hrule height 1pt}
    $\ket{1}$ & $\ket{2}$ & $\ket{3}$ & $\ket{4}$ \tabularnewline
    \hline
    $\ket{3}$ & $\ket{4}$ & $\ket{a_{12}}$ & $\ket{b_{12}}$ \tabularnewline
    \noalign{\hrule height 1pt}
    $\ket{2}$ & $\ket{1}$ & $\ket{4}$ & $\ket{3}$ \tabularnewline
    \hline
    $\ket{4}$ & $\ket{3}$ & $\ket{b_{12}}$ & $\ket{a_{12}}$ \tabularnewline
    \noalign{\hrule height 1pt}
    \end{tabular}\ .
\end{equation}
It is noticeable that even if we give all the computational vectors from the above grid as clues, we still have freedom in choice of $\ket{a_{12}}$ and $\ket{b_{12}}$. In particular, we can set $\ket{a_{12}}=\ket{1}$, $\ket{b_{12}}=\ket{2}$ or vice versa. Hence, the original classical grid was not uniquely solvable.

Finally, SudoQ with the cardinality $c=4$, whose vectors are chosen in the computational basis corresponds to the classical solution, which we assumed is unique.

\section{Distinct eigenvectors of Weyl-Heisenberg operators $XZ^k$ and $XZ^l$}\label{sec:WH_app}
Let $X$ and $Z$ be standard shift and phase operators defined on the $N$-dimensional Hilbert space $\mathcal{H}_N$
\begin{equation}
X= \sum_{j=1}^N \ket{j}\bra{j+1}, \quad Z=  \sum_{j=1}^N \omega^{j}  \ket{j}\bra{j}.
\end{equation}

Notice that the following commutation relation holds:
\begin{equation}
    XZ = \omega ZX,
\end{equation}
where $\omega = e^{2\pi i/ N}$ is a root of unity. Consequently, $XZ^k = \omega^k ZX$.
\begin{proposition}
No eigenvector of $XZ^k$ is an eigenvector of $XZ^l$ for $k\neq l$.
\end{proposition}
\begin{proof}
Suppose that $\ket{v}$ is a simultaneous eigenvector of $XZ^k$ and $XZ^l$
\begin{equation}
    XZ^k \ket{v} = \lambda_k \ket{v}, \qquad XZ^l \ket{v} = \lambda_l \ket{v}.
\end{equation}
Without loss of generality, we can assume $k>l$, which enables us to transform above relationship into
\begin{equation}
    XZ^k \ket{v} = XZ^{k-l}Z^l \ket{v} = \omega^{k-l}\lambda_l Z^{k-l} \ket{v}.
\end{equation}
Therefore, $\ket{v}$ is also an eigenvector of $Z^{k-l}$.
If we expand $\ket{v}$ in the computational basis $\ket{v} = \sum_i \alpha_i \ket{i}$, we can note that there exists $i$ such that $\alpha_i$ equals zero. This is true if the dimension $N$ is prime, as $Z^{k-l}$ has a non-degenerated spectrum and thus its eigenvectors are elements of the computational basis $\mathcal{B}$, with only one non-zero coefficient.
However, even if the spectrum of $Z^{k-l}$ is degenerated, it is not fully degenerated, so combining it with the diagonal form of this operator we obtain that indeed some coefficient $\alpha_i$ is zero.

On the contrary, for an eigenvector $\ket{v} = \sum_i \alpha_i \ket{i}$ of the matrix $XZ^k$ all coefficients $\alpha_i$ are non-zero since the action of the matrix on a vector translates all its elements by one, possibly with some aquired phase.

Gathering the above statements together, by contradiction we conclude that no vector $\ket{v}$ is a simultaneous eigenvector of $XZ^k$ and $XZ^l$.
\end{proof}

\bibliography{Bib}
\end{document}